\documentclass[useAMS,usenatbib]{mn2e}
\usepackage{graphicx}
\usepackage{txfonts}
\title{On the inference of stellar ages and convective-core properties in main-sequence solar-like pulsators}
\author[I.\,M.\,Brand\~ao,  M.\,S.\,Cunha and J.\,Christensen-Dalsgaard]{I.\,M.\,Brand\~ao$^{1}$\thanks{E-mail:isa@astro.up.pt},
 M.\,S.\,Cunha$^{1}$ and J.\,Christensen-Dalsgaard$^{2}$\\
$^{1}$Centro de Astrof\'isica and Faculdade de Ci\^encias, Universidade do Porto, Portugal, Rua das Estrelas, 4150-762 Porto, Portugal\\
$^{2}$Stellar Astrophysics Centre, Department of Physics and Astronomy, Aarhus University, Ny Munkegade 120, DK-8000 Aarhus C, Denmark}
\begin{document}
\date{Accepted 2013 November 29.  Received 2013 November 5; in original form 2013 July 18}
\pagerange{\pageref{firstpage}--\pageref{lastpage}} \pubyear{2013}
\maketitle
\label{firstpage}
\begin{abstract}
Particular diagnostic tools may isolate the signature left on the oscillation frequencies by the presence of a small convective core. 
Their frequency derivative is expected to provide information about convective core's properties and stellar age. 
The main goal of this work is to study the potential of the diagnostic tools with regard to the inference of stellar age and stellar core's properties. 
For that, we computed diagnostic tools and their frequency derivatives from the oscillation frequencies of 
main-sequence models with masses between 1.0 and $1.6\,{\rm\,M}_{\sun}$ and with different physics. 
We considered the dependence of the diagnostic tools on stellar age and on the size of the relative 
discontinuity in the squared sound speed at the edge of the convectively unstable region. We find that the 
absolute value of the frequency derivatives of the diagnostic tools increases as the star evolves on the main sequence. 
The fraction of stellar main-sequence evolution for models with masses $>1.2\,{\rm\,M}_{\sun}$ may be estimated from the 
frequency derivatives of two of the diagnostic tools. For lower mass models, constraints on the convective core's overshoot can 
potentially be derived based on the analysis of the same derivatives. For at least 35 per cent of 
our sample of stellar models, the frequency derivative of the diagnostic tools takes its maximum absolute 
value on the frequency range where observed oscillations may be expected.
\end{abstract}
\section{Introduction}
\label{sec1}
According to stellar structure theory, 
stars slightly more massive than the Sun ($M > 1.0\,{\rm\, M}_{\sun}$) may develop a 
convective core at some stage of their evolution on the main sequence.
This occurs when radiation can no longer transport the amount of energy
produced by nuclear burning in the internal regions of a star. For these stars
hydrogen burning happens primarily through the CNO cycle, which has a strong
dependence on the temperature. Convection is then the most effective process
to transport the energy, hence a convective core is present.
Since convection implies chemical mixing, the 
evolution of these stars is severely influenced by the presence
and the extent of the convective core.

Models of an intermediate-mass star 
($1\,{\rm\,M}_{\sun} < M \leq 2\,{\rm\,M}_{\sun}$)
with a convective core show that the extent in mass 
of the convective core does not remain constant 
during the evolution of the star. It increases during the initial stages of
evolution before beginning to decrease, later on. 
Also, some physical ingredients such as the convective-core overshoot, 
if present, will influence the extent and evolution of the convective core. 

A convective core is believed to be homogeneously mixed since the time-scale 
for mixing of elements is much shorter than the nuclear time-scale. Thus,  
if diffusion is not taken into account, the growing
core causes a discontinuity in the composition at the edge of the core \citep{mitalas72,saio75},
with a consequent density discontinuity. 
If diffusion is present, instead of a discontinuity, there will be a very steep gradient 
in the chemical abundance and density at the edge of the convective core.
Moreover, a retreating core leaves behind a non-uniform chemical profile \citep{faulkner73}
causing also a very steep gradient in the chemical abundance.  

This paper concerns the study of the properties of the convective cores
in main-sequence models of solar-like pulsators. One of the main goals 
of inferring information of the deepest layers of
stars is to improve the description
of particular physical processes such as diffusion and convective overshoot, 
in stellar evolution codes. That, in turn, will improve
the mass and age determinations derived from asteroseismic
studies.
This work presented here is driven,
in particular, by the following questions: can we detect the signature of a small 
convective core on the oscillation frequencies of solar-like pulsators for which 
photometric data with the quality such as that of the NASA \textit{Kepler} satellite \citep{borucki10} exist?
What is the dependence of this signature on the stellar mass and physical parameters?
What is the precision required on the individual frequencies in order to detect the signature of 
a convective core? Will the detection of such a signature provide information
about the stellar age?

To address these questions, we will focus on the analysis of 
a number of seismic diagnostic tools
presented in Section \ref{sec2}. These diagnostic tools
were computed for a set of main-sequence solar-like
models through a method described in Section \ref{sec3}.
In Section \ref{sec4}, we present our results and in
Section \ref{sec5}, we conclude.
\section{Diagnostic tools}
\label{sec2}
\begin{figure}
\begin{center}
\includegraphics[width=9cm,angle=0]{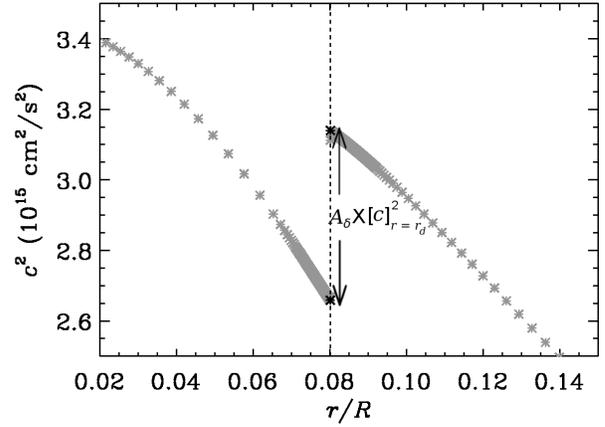}
\end{center}
\caption{The square of the sound speed in the innermost 
regions of a $1.4\,\rm M_{\sun}$ model with an age of 1.56 Gyr
computed with solar metallicity and assuming an overshoot from the
convective core of $\alpha_{\rm OV}= 0.1$. 
The dashed vertical line corresponds to the location
of the discontinuity where the sound-speed derivative is maximum. 
The two black star symbols mark the position of the two extremes of the discontinuity, 
that are used to compute its amplitude
in an automated manner.}
\label{fig:c2_rr}
\end{figure}

The oscillation frequencies of a pulsating star depend not only on its global properties, such
as the mass and radius, but also on the details of its internal structure, 
including any  sharp structural variations, such as those that are present
at the borders of convectively unstable regions. 
Different combinations of low-degree p modes have been
shown to probe the interior of stars \citep[e.g.][]{cd84}.
Among these, the most commonly used are the large frequency
separation  defined as the difference in frequency of modes of the same degree, $l$, and 
consecutive radial order, $n$, $\Delta \nu_{n,l} = \nu_{n+1,l} - \nu_{n,l}$ 
\citep[e.g.,][]{cd79a,cd79b}, and
the small frequency separation between
modes of degrees $l = 0$ and 2,
$d_{02}  = \nu_{n,0} - \nu_{n-1,2}$
\citep[e.g.][]{gough83}. In addition, \cite{roxburgh03}
proposed smooth five-points small frequency separations,
$d_{01}$ and $d_{10}$, as a diagnostic of stellar interiors.
These small separations are defined by
\begin{equation}
\label{eq:d01}
d_{01}(n)=\frac{1}{8}(\nu_{n-1,0}-4\nu_{n-1,1}+6\nu_{n,0}-4\nu_{n,1}+\nu_{n+1,0})
\end{equation}
\begin{equation}
\label{eq:d10}
d_{10}(n)=-\frac{1}{8}(\nu_{n-1,1}-4\nu_{n,0}+6\nu_{n,1}-4\nu_{n+1,0}+\nu_{n+1,1}).
\end{equation}
All these diagnostic tools are, however, affected by the poorly modelled 
outer layers of stars.
Nevertheless, \cite{roxburgh03} demonstrated that the effect of the outer stellar regions
in the oscillation frequencies is cancelled out when one considers
the ratios of these diagnostics
\citep[see also][]{roxburgh04,oti05,roxburgh05}. 
These ratios are defined by
\begin{eqnarray}
\label{eq:rat}
r_{01}(n)=\frac{d_{01}(n)}{\Delta \nu_{n,1}},& & r_{10} = \frac{d_{10}(n)}{\Delta \nu_{n+1,0}}\,,
\end{eqnarray}
\begin{eqnarray}
\label{eq:r02}
r_{02}(n)=\frac{d_{02}(n)}{\Delta \nu_{n,1}}\,,
\end{eqnarray}
and $d_{02}$ is the small frequency separation previously defined.
While $d_{02}$, $d_{01}$, $d_{10}$ and the corresponding ratios probe the inner regions of the star, 
they do not isolate the signature left of the oscillation frequencies by the 
region of rapid structural variation associated with the edge of a convective core. 
An illustration of that structural variation is shown in Fig. \ref{fig:c2_rr}, in terms of its 
impact in the sound-speed profile. The functional form of the corresponding signature 
on the frequencies was derived  by \cite{cunhamet07} based on the properties 
of the oscillations of stellar models slightly more massive than the Sun. 
The authors assumed that this signature is caused by the discontinuity in the composition
and hence in the sound speed at the edge of the growing convective core.
They showed that the following combination of oscillation frequencies
is sensitive to the properties of the sound-speed discontinuity, and
is also capable of isolating the consequent perturbation to the oscillation frequencies:
\begin{equation}
\label{eq:dr0213}
dr_{0213} = 6 \left( \frac{D_{02}}{\Delta \nu_{n-1,1}} - \frac{D_{13}}{\Delta \nu_{n,0}}\right).
\end{equation}
This diagnostic tool corresponds to a 
difference of ratios
between the scaled small separations,
$D_{l,l+2}\equiv(\nu_{n,l}-\nu_{n-1,l+2})/(4l+6)$, and the
large separations, $\Delta\nu_{n,l}$,
for different combinations of mode degrees. Note that
the authors originally defined this diagnostic tool without
the factor 6. However, since the relation between
the perturbation and the scaled small separations has a factor 6,
we opted to include it here in the definition of $dr_{0213}$.
More recently, \cite{cunha11} improved the analysis
presented in \cite{cunhamet07}, by considering 
a different expression to describe the sound-speed variation
at the edge of the growing convective core, which is more in line
with the variation observed from the equilibrium models. Moreover,
they showed that the frequency derivative of the diagnostic tool $dr_{0213}$
can potentially be used to infer the amplitude of the relative sound-speed variation
at the edge of the growing convective core, $A_\delta \equiv [\delta c^2/c^2]_{r=r_{\rm d}}$, with $r_{\rm d}$ 
being the radial position at which the discontinuity
in the sound speed occurs (cf. Fig. \ref{fig:c2_rr}), and $c^2$ at $r=r_{\rm d}$
is the highest of the two $c^2$ values at the discontinuity.
The disadvantage of this diagnostic tool is that it requires the  observation of modes of degree up to 3.  

In this work, we study the behaviour of the 
diagnostic tools defined by Eqs.\,(\ref{eq:rat})-(\ref{eq:dr0213}).
For that, we derived these tools for a set of stellar models
considering a large parameter space.
We then compare the frequency behaviour of the three diagnostic tools 
and consider their potential concerning 
the inference about the properties of stellar cores and the inference of stellar ages. 
\section{Method}
\label{sec3}
\subsection{Grids of models}
We started by computing a set of evolutionary tracks
with the \lq Aarhus STellar Evolution Code\rq\, 
\citep[\textsc{astec};][]{cd08aastec}. The models were
calculated using the most up-to-date \textsc{opal} 2005
equation-of-state tables\footnote{\textsc{opal} tables available at 
{\it $http\colon//opalopacity.llnl.gov/EOS\_2005/$}.}
\citep{rogers02}, \textsc{opal}95 opacities \citep{iglesias96}
complemented by low-temperature opacities from \cite{ferguson05}.
The opacities were calculated with the solar mixture
of \cite{grevesse93}, and the Nuclear Astrophysics Compilation of REaction Rates
\citep[NACRE;][]{angulo99}.
Convection was treated according the standard mixing-length theory
from \cite{vitense58}, where the characteristic length
of turbulence, called the mixing length $l_{\rm{ML}}$, 
scales directly with the local pressure scale-height,
$H_{\rm p}$, as $l_{\rm{ML}} = \alpha_{\rm ML} H_{\rm p}$,
with $\alpha_{\rm ML} > 1$. 
Overshooting (OV) exists when the convective movements
of the gas in the convectively unstable regions
cause extra mixing beyond the border of such regions.
In our work, we only considered convective-core overshoot.
In practice, since the core can be very small, \textsc{astec} assumes
$\alpha_{\rm OV} \min(r_{\rm core}, H_{\rm p})$, where 
$r_{\rm core}$ is the radius of the convective core.
Regarding the atmosphere, we considered an atmospheric temperature versus optical
depth relation which is a fit to the quiet-sun relation 
of \cite{vernazza76}. Diffusion and settling were not taken into account.

\begin{table}
\centering
\small
\caption{Parameters used to compute the evolutionary tracks. $M/\rm{M_{\sun}}$ is
the mass in solar units, $Z/X$ is the initial ratio of heavy elements to hydrogen abundances and
$Y$ the helium abundance. $\alpha_{\rm ML}$ is the mixing-length parameter and
$\alpha_{\rm OV}$ the core overshoot parameter. We considered $(Z/X)_{\sun} = 0.0245$ for the Sun \citep{grevesse93}.}
\label{tab:inputp}
\begin{tabular}{@{}lll}
\hline\hline
Parameter & Grid I & Grid II\\
\hline
$M/\rm{M_{\sun}}$ & 1.0-1.6 (0.1 steps)& 1.0-1.6 (0.1 steps)\\
$Z/X$ & 0.0245 & 0.0079, 0.0787\\
$Y$ & 0.278 & 0.255, 0.340\\
$\alpha_{\rm ML}$& 1.8 & 1.8\\
$\alpha_{\rm OV}$& 0.0-0.2 (0.1 steps) & 0.1\\
\hline
\end{tabular}
\normalsize
\end{table}

All of the evolutionary tracks computed contain a fixed number of models, 
not equally spaced in time, from the zero-age main sequence (ZAMS) to the post-main sequence.
The parameter space that we considered in the modelling is shown 
in Table\,\ref{tab:inputp}.

We constructed two grids of evolutionary tracks, Grid\,I and Grid\,II. 
For the former, we considered solar metallicity, i.e. [Fe/H] = 0, where
$\rm (Z/X)_{\sun} = 0.0245$ for the Sun \citep{grevesse93}, and varied $\alpha_{\rm OV}$,
while in Grid\,II, we fixed $\alpha_{\rm OV} = 0.1$ and considered two extreme values 
for the metallicity, namely [Fe/H] = -0.5 and [Fe/H] = 0.5. To convert
from [Fe/H] to the initial ratio of heavy elements to hydrogen abundances, $Z/X$, we used the relation
$\textrm{[Fe/H]} = \log \left( Z/X \right)- \log \left(\rm Z/X \right)_{\sun}$,
where $\rm{[Fe/H]}$ is the star's metallicity, $Z$
and $X$ are all elements heavier than helium and hydrogen mass fractions, respectively, and
$(Z/X)_{\sun}$ is the ratio for the solar mixture.
The value of helium, $Y$, was obtained from the relation 
$Y = Y_{\rm p} + Z\,\textrm{d}Y/\textrm{d}Z$,
where $Y_{\rm p}$ is the abundance of helium
produced during primordial nucleosynthesis
and $\textrm{d}Y/\textrm{d}Z$ is the helium
to metal enrichment ratio.
Considering $\textrm{d}Y/\textrm{d}Z=2$ \citep[see, e.g.,][]{casagrande07}
and using the solar values of $\rm (Z/X)_{\sun} = 0.0245$ \citep{grevesse93} and $\rm Y_{\sun} = 0.278$
\citep{serenelli10} we found $Y_{\rm p} = 0.2435$.
Using $\textrm{d}Y/\textrm{d}Z=2$ and $Y_{\rm p} = 0.2435$,
and fixing $Z$, we derive $Y$. 

We focused our work on intermediate-mass models, 
i.e. $1.0\,{\rm\,M_{\sun}}\leq\,M\,\leq 1.6\,{\rm\,M_{\sun}}$, because 
it is in this mass range that we expect main-sequence stars to show
solar-like pulsations.
We fixed the mixing-length parameter to 1.8. We note that changing the mixing length parameter 
does not affect the convective core. In this region, the temperature gradient 
takes approximately its adiabatic value and is,
therefore, independent of the details of the theory of convection. 
The values for the metallicity were chosen so that they comprise
those derived for the solar-like stars observed by the \textit{Kepler} satellite.
The value of the convective overshoot for stars with masses
$M < 1.7\,\rm M_{\sun}$ is quite unknown with literature
values of $\alpha_{\rm OV}$ ranging from 0.00 to 0.25 \citep{ribas00}.
We note that recent modellling of the \textit{Kepler} main-sequence solar-like pulsator KIC\,1200950 (also
known as Dushera within the Kepler Asteroseismic Science Consortium), with an inferred mass 
of around 1.5\,M$_{\sun}$, required the presence
of mixing beyond the boundary of its formal convective core \citep{aguirre13}.
We considered values for the convective-core overshoot between 0.0 and 0.2.
Since our goal was only to study the
dependence of the diagnostic tools on the input parameters,
we did not construct refined grids. 

\subsection{Models selection}
\begin{figure}
\begin{center}
\includegraphics[width=9cm,angle=0]{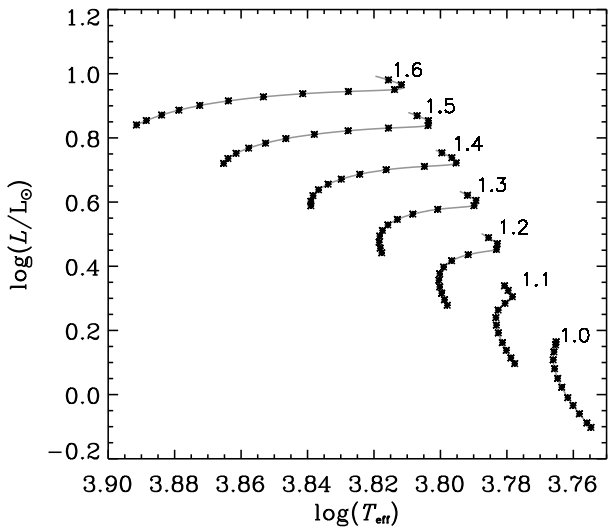}
\end{center}
\caption{The HR diagram for a set of models with solar metallicity,
$Z/X=0.0245$, and no convective-core overshoot. The models were
evolved from the ZAMS to the post-main sequence.  Each 
evolutionary track contains 300 models at different evolutionary stages.
Since the number of the computed models is fixed and the time step varies
with mass and also along the evolution, 
the exact position of the 
last model of each evolutionary track depends on its mass.
The numbers at the end of each evolutionary track correspond
to the mass, in solar units. The 12 selected models within each evolutionary track
are shown by a star symbol (see the text for details).}
\label{fig:ms12}
\end{figure}

Each evolutionary track contains up to 
200 models within the main-sequence phase, defined here as models
with hydrogen abundance in the core $X_{\rm c} > 10^{-2}$. 
The true number of models depends mainly on 
the mass associated with the track. 
Rather than analysing all the $\lesssim200$ models,
we considered 12 models
along the main-sequence phase
that are equally spaced in $\log\,g$, $\log T_{\rm eff}$ and $\log (L/\rm L_{\sun})$.
To select the 12 models, we started by computing the 
total parameter distance, $d_{\rm tot,param}$, 
that a given model travels along
the main sequence in a 3D space
with the following parameters: $\log\,g$, $\log\,T_{\rm eff}$ and $\log (L/\rm\,L_{\sun})$.
This distance is defined by
\begin{equation}
 d_{\rm tot,param} = \sum_{i=1}^{N-1} d_{i,\rm param},
\end{equation}
and,
\begin{eqnarray}
 d_{\rm i,param} & =& \big( \left[\log\,g(i+1)-\log\,g(i)\right]^2 \nonumber\\
&& + \left[\log\,T_{\rm eff}(i+1)- \log\,T_{\rm eff}(i)\right]^2 \\
&& + \left[\log\,(L/{\rm L_{\sun}})(i+1)-\log\,(L/{\rm L_{\sun}})(i)\right]^2\big)^{1/2}, \nonumber
\end{eqnarray}
where $N$ is the total number of main-sequence models in each evolutionary track.
The total distance was equally divided in 12 segments, and the
12 models were chosen such as to have their $\log\,g$, $\log T_{\rm eff}$, and $\log (L/\rm\,L_{\sun})$
the closest to the respective values for the 12 segments.
Fig. \ref{fig:ms12} shows the Hertzsprung-Russell (HR) diagram for 
a set of main-sequence models with mass varying
between $1.0$ and $1.6\,\rm\,M_{\sun}$, with solar metallicity
and no convective-core overshoot. The 12 models are represented
by the black star symbol. In this plot, the models shown are not equally
spaced since, for clarity, the third dimension, namely $\log\,g$, is not represented.

For the 12 models within each track we computed oscillation frequencies 
using the Aarhus adiabatic oscillation code \citep[\textsc{adipls};][]{cd08adipls}.
\subsection{Diagnostic tools from models}
In the case of stars with convective cores, we expect that the frequency 
derivatives of the diagnostic tools taken at their maximum absolute value 
are a measure of the amplitude of the discontinuity 
in the sound speed at the edge of the growing convective core \citep{cunha11}
and, hence, that they are strongly sensitive to the stellar age. 
\begin{figure}
\begin{center}
\includegraphics[width=9cm,angle=0]{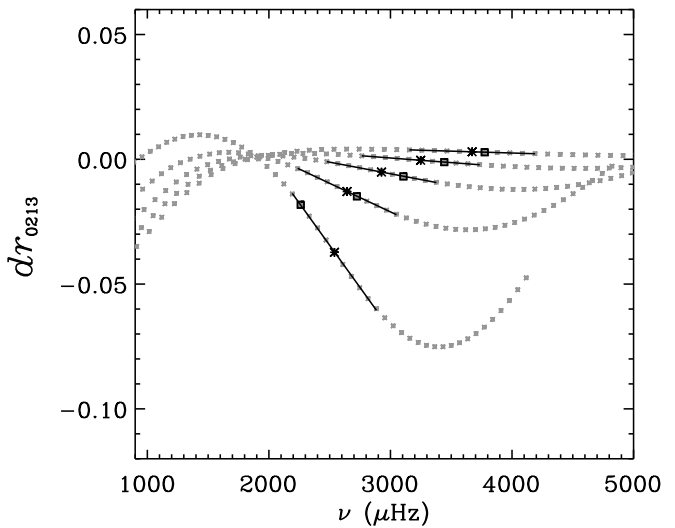}
\end{center}
\caption{The diagnostic tool $dr_{0213}$
as a function of frequency, $\nu$, for 
a sequence of $1.3\,\rm\,M_{\sun}$ models with solar metallicity and without core 
overshoot. The different curves correspond to models with different ages, 
the most evolved model being the one with the largest absolute value of the derivative.
The dark stars represent the frequencies at which the derivative takes its largest absolute value, 
for each model, and the straight black lines that cross that minima correspond to 
the region of approximately constant derivative (where the slopes are computed). 
The squares represent the acoustic cut-off frequency, $\nu_{\rm c}$, computed for each model. Note that for
the oldest model, the region of constant derivative is above
the cut-off frequency for the physics considered in our models.}
\label{fig:dr0213_awcl}
\end{figure}
An example of how the frequency derivative of $dr_{0213}$ 
at its maximum absolute value
changes during the main-sequence evolution is shown in Fig. \ref{fig:dr0213_awcl}.  
It is quite clear that the absolute value of the derivative increases with age. Nevertheless, 
it is also clear from inspection of the figure that the frequency 
at which the absolute value of the derivative reaches its maximum
is not always below the acoustic cut-off frequency, $\nu_{\rm c}$, 
i.e., below the maximum frequency such that acoustic waves are expected to be 
contained within the star). Consequently, some limitations are to be expected 
when trying to use the frequency derivatives of the seismic diagnostics tools 
as a way for determining stellar ages.

With the above in mind, 
we shall consider two different tasks in our study. First, we shall verify how strongly 
related are the frequency derivatives of the diagnostic tools taken at their maximum absolute value
and the amplitude of the discontinuity  
in the sound speed at the edge of the convective core, using all models in our grid. 
For that, we will require that the oscillation frequencies in each model cover the 
region where the maximum absolute value of the derivatives is placed. 
Consequently, we apply a full reflective outer boundary condition in \textsc{adipls},
expressed by $\delta p = 0$, where $\delta p$ is the Lagrangian
perturbation to the pressure, to obtain eigenfrequencies 
above the acoustic cut-off frequency. Later, we consider 
the observational impact of our results, by analysing the subset of models for 
which the maximum absolute value of the derivatives is below the acoustic cut-off frequency.
\begin{figure}
\begin{center}
\includegraphics[width=9cm,angle=0]{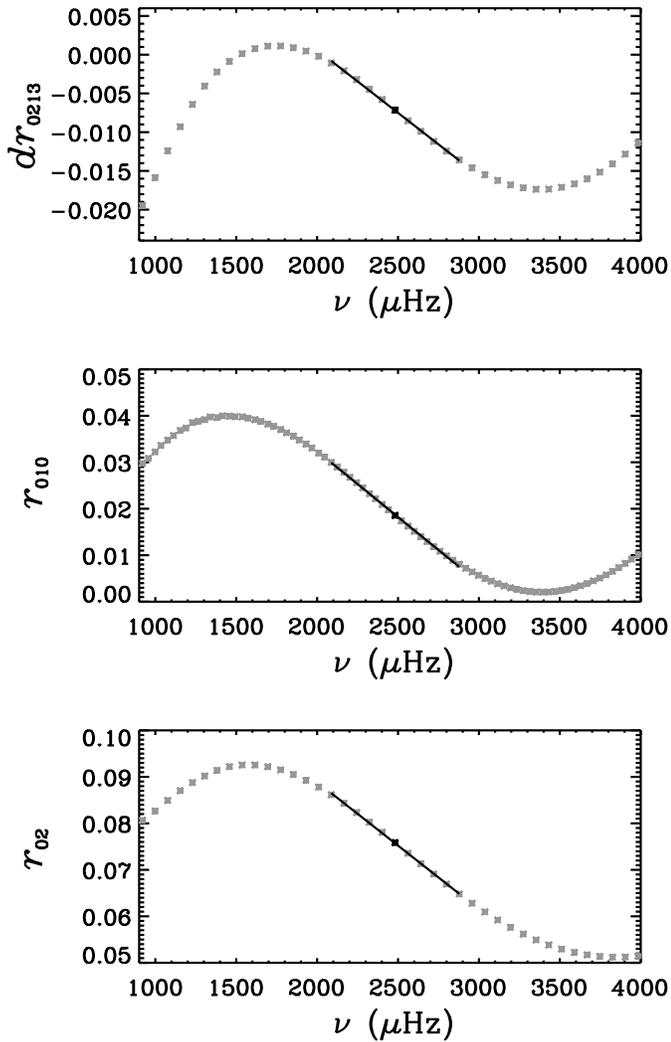}
\end{center}
\caption{The three diagnostic tools,
namely $dr_{0213}$ (upper panel), $r_{010}$ (middle panel) and
$r_{02}$ (lower panel) as a function of frequency, computed 
for a $1.4\,\rm\,M_{\sun}$ model
without core overshoot and with solar metallicity. This model
has an age of 1.29\,Gyr and corresponds to the sixth model
of the $1.4\,\rm\,M_{\sun}$ evolutionary track of Fig. \ref{fig:ms12}.}
\label{fig:slope_all_min}
\end{figure}
\begin{figure}
\begin{center}
\begin{minipage}[t]{\linewidth} 
\centering
\includegraphics[width=9cm,angle=0]{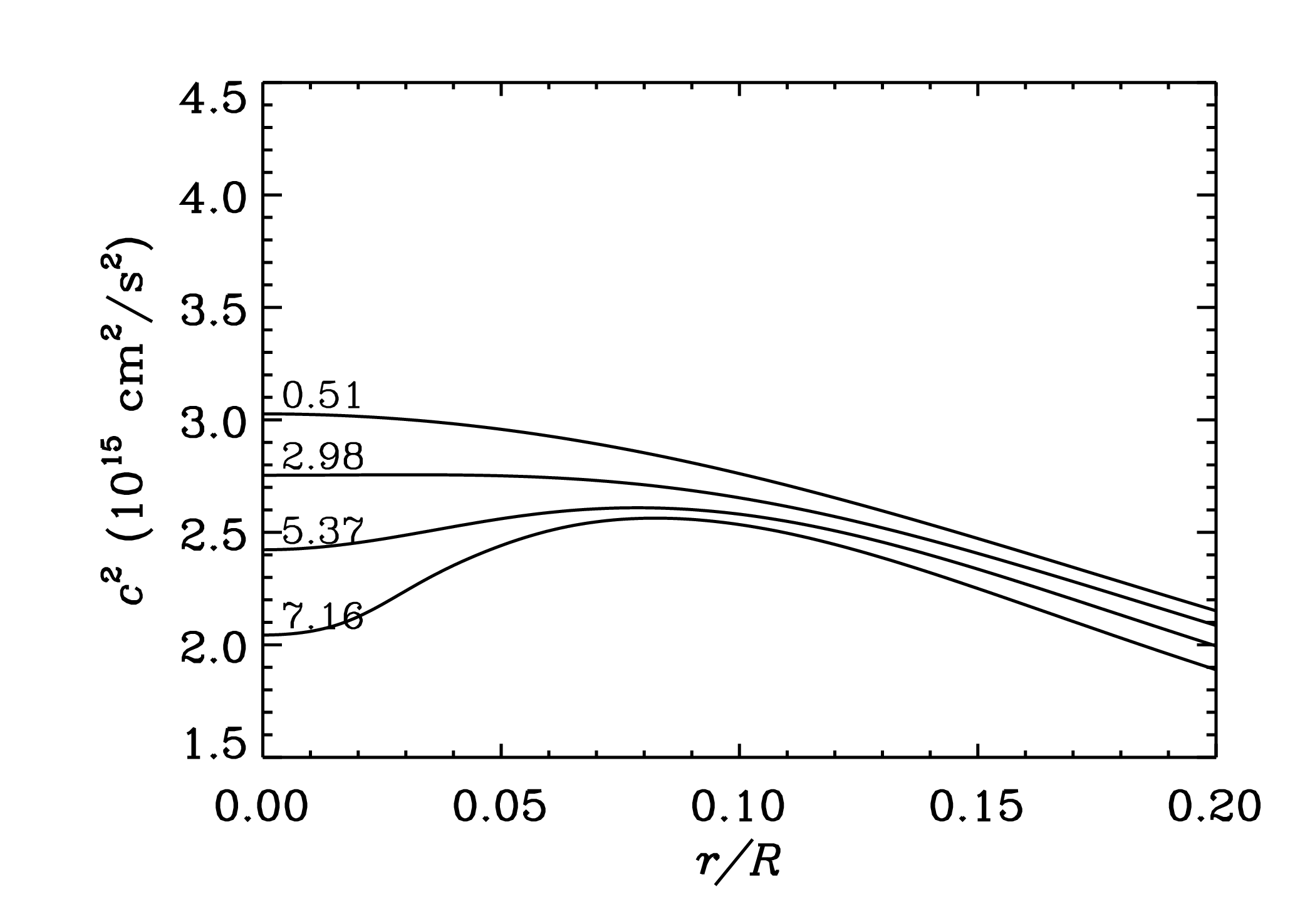} \\
\includegraphics[width=9cm,angle=0]{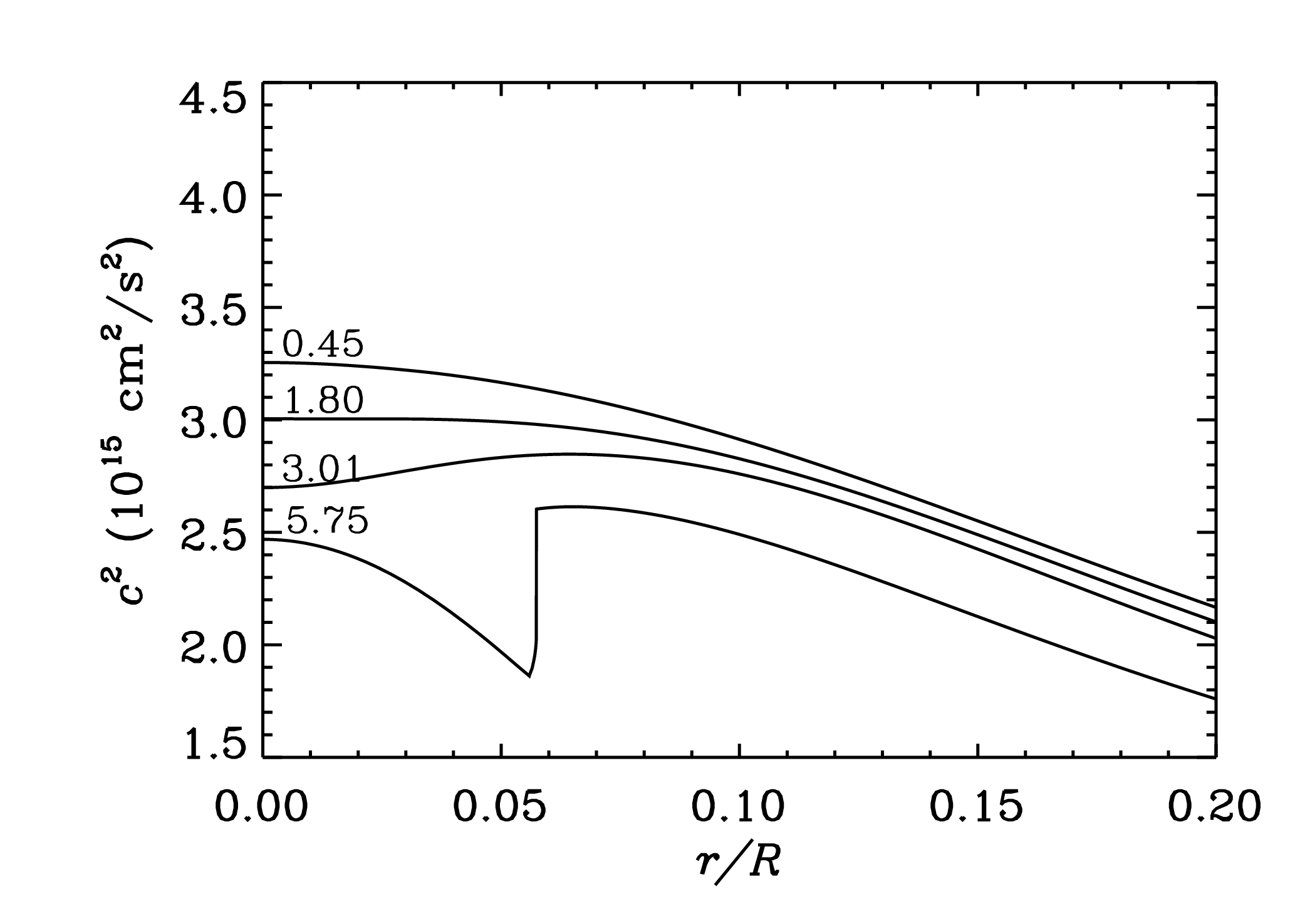}\\
\includegraphics[width=9cm,angle=0]{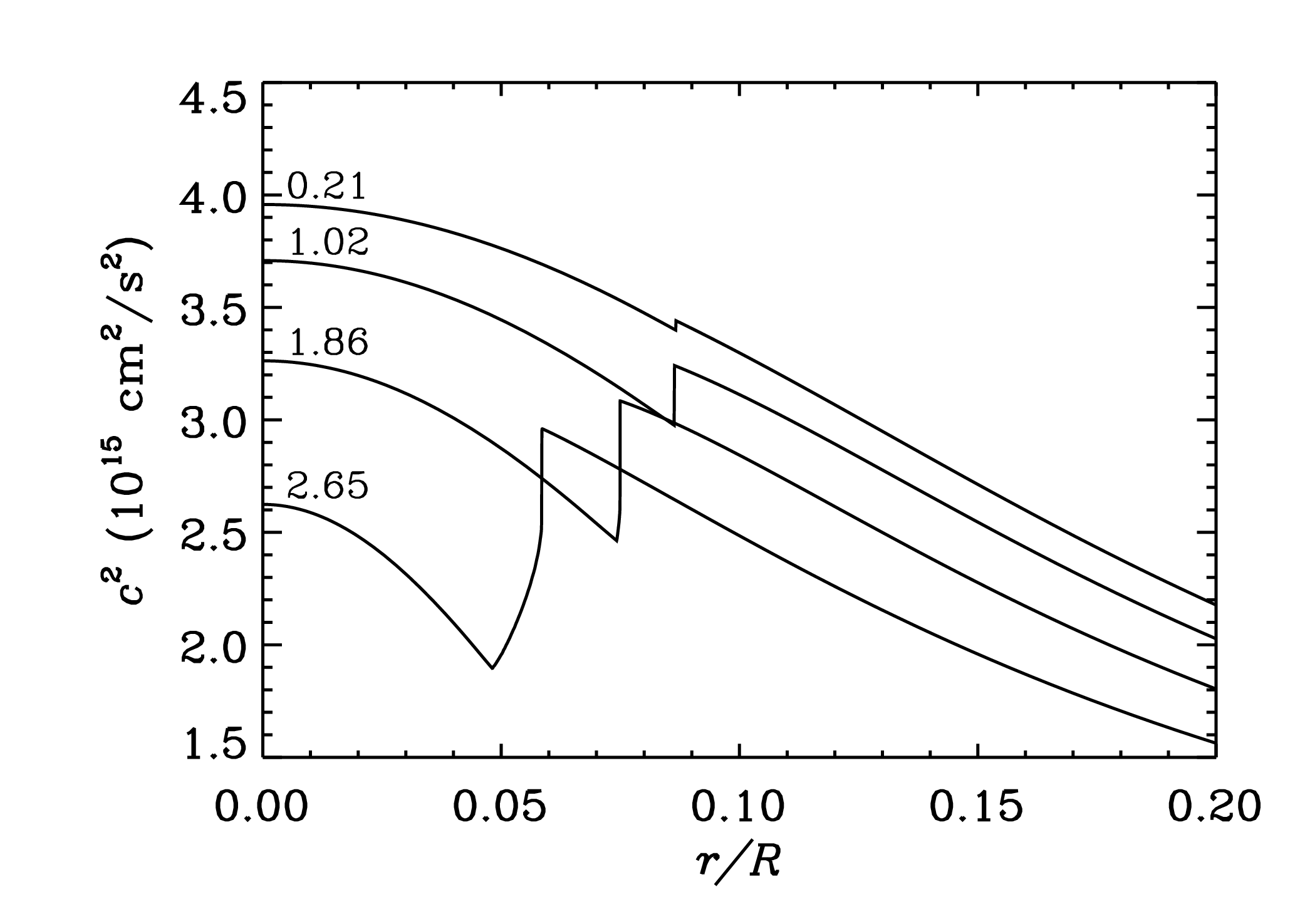}\\
\end{minipage}
\end{center}
\caption{The sound-speed profile in the
inner layers of a $M=1.0\,\rm M_{\sun}$ (upper panel), $M=1.1\,\rm\,M_{\sun}$
(middle panel) and $M=1.4\,\rm M_{\sun}$ (lower panel) with different ages.
The numbers at the beginning of the curves indicate the age of the models 
in Gyr. These models were computed with 
solar metallicity and assuming overshoot 
from the convective core of $\alpha_{\rm OV} = 0.1$.}
\label{fig:c2_rr_m}
\end{figure}

For practical purposes we introduce here a new quantity $S\{i\}$, which we shall name \lq slope\rq, 
which is a measure of the frequency derivative of a diagnostic tool \lq $i$\rq, at its maximum absolute value. 
To calculate the slopes of the diagnostic tools we proceed as follows:
we start by inspecting the behaviour of the diagnostic tool $dr_{0213}$ as a function
of frequency to determine the frequency  $\nu_{\rm slope}$ at which the maximum
absolute value of the derivative of $dr_{0213}$ is placed
and identify the corresponding radial order, $n_{\rm slope}$. 
We then perform a linear least-squares fit to 
the 10 frequencies of modes of consecutive radial orders, $n$, centred on $n_{\rm slope}$.
The slope of the diagnostic tool $dr_{0213}$
around $\nu_{\rm slope}$ is given by the linear coefficient obtained from this linear fit.
The slopes of the diagnostic tools $d_{01}$, $d_{10}$ and $d_{02}$, and their
respective ratios, are computed in the same range
of frequency as considered for $dr_{0213}$. 
Note that in relation to the
quantities $d_{01}$ and $d_{10}$, they are both
five-point combination of modes of degrees $l=0$ and 1, the first
centred on modes of degree $l=0$ and the second
on modes of degree $l=1$. In practice, we consider
these two quantities together, denoting the concatenation of the two,
appropriately ordered in frequency,
by $d_{010}$, and the corresponding ratios by $r_{010}$.
For this quantity, the slope is also determined
in the same frequency range as that considered for 
$dr_{0213}$, but instead of 10, 20 frequencies are used.
In Fig. \ref{fig:slope_all_min},
we show as an example, the diagnostic tools $dr_{0213}$, $r_{010}$ and
$r_{02}$ computed for a $1.4\,\rm M_{\sun}$ model
without overshoot and with solar metallicity at an 
age of 1.29\,Gyr, and the frequency region considered
in the computation of the slopes.

To have an estimation of the error associated with
the slope computed for each diagnostic tool, we 
chose seven models with different values for the mass and input
physics, and with different values of $A_\delta$. For these
models, we randomly generated 10\,000 sets of model frequencies 
within the error, assuming a relative error of $10^{-4}$
for each individual frequency. For each generation,
we computed the slopes in the same manner as described above.
We then computed the mean of the 10\,000 values obtained for the slopes
and the standard deviation was considered to be our error
estimation for the slopes of the diagnostic tools.

As mentioned above, models that have a convective core
show a structural discontinuity at the edge of the 
convectively unstable region. Here, we aim at 
studying the relation between the relative amplitude $A_\delta$ of the
discontinuity in the squared sound speed and the
slopes of the diagnostic tools, $S\{i\}$. 
Thus, for the models for which 
we computed the diagnostic tools
and that have a convective core, 
we also computed $A_\delta$. To compute
the latter, we started
by identifying the location (in terms of $r/R$)
of the discontinuity in $c^2$. To do so, we 
identified the maximum value of the derivative
of the squared sound speed in the inner
regions of the models. This maximum value was used to determine
the precise location of the discontinuity in an automated manner, for all models. 
With that we were able to automatically compute the 
actual amplitude of the sound-speed discontinuity by measuring the difference between the maximum
and minimum values of $c^2$ at the discontinuity. These two values are shown, as an example,
by the two black stars in Fig. \ref{fig:c2_rr}.
\section{Results}
\label{sec4}
\subsection{Models with a convective core}
\label{sec4.1}
Since our present study concerns only models with a convective core, 
we started by identifying which models in our grid are included in this category. 
We verified that no convective core exists in our 
$1.0\, {\rm M}_{\sun}$ sequences of main-sequence models.
In Fig. \ref{fig:c2_rr_m}, upper panel, 
we show an example of the sound-speed profile in the innermost regions 
of a sequence of $1.0\,\rm M_{\sun}$ models with solar metallicity.
Clearly, these models do not show a sound-speed discontinuity, or even 
strong sound-speed gradients in the innermost layers, 
although an increase in the sound-speed gradient is seen as 
the model star evolves towards the terminal-age main sequence (TAMS).
On the other hand, the presence or absence of a convective core in models with
$1.1\,{\rm M}_{\sun}$ depends on the metallicity considered. 
There are no models with a convective core in our lowest metallicity sequences of $1.1\,{\rm M}_{\sun}$, 
namely with $Z/X=0.0079$. 
The most evolved models with $1.1\,{\rm M}_{\sun}$ and with solar
$Z/X=0.0245$ metallicity show a convective core (Fig. \ref{fig:c2_rr_m}, middle panel) and 
the $1.1\, {\rm M}_{\sun}$ models with high metallicity, $Z/X=0.0787$,
all have a convective core.
Models with $M \geq 1.2\,{\rm M}_{\sun}$ all have convective cores           
(Fig. \ref{fig:c2_rr_m}, lower panel). 
\begin{figure*}
\centering
\begin{minipage}{\linewidth}
\includegraphics[width=9cm,angle=0]{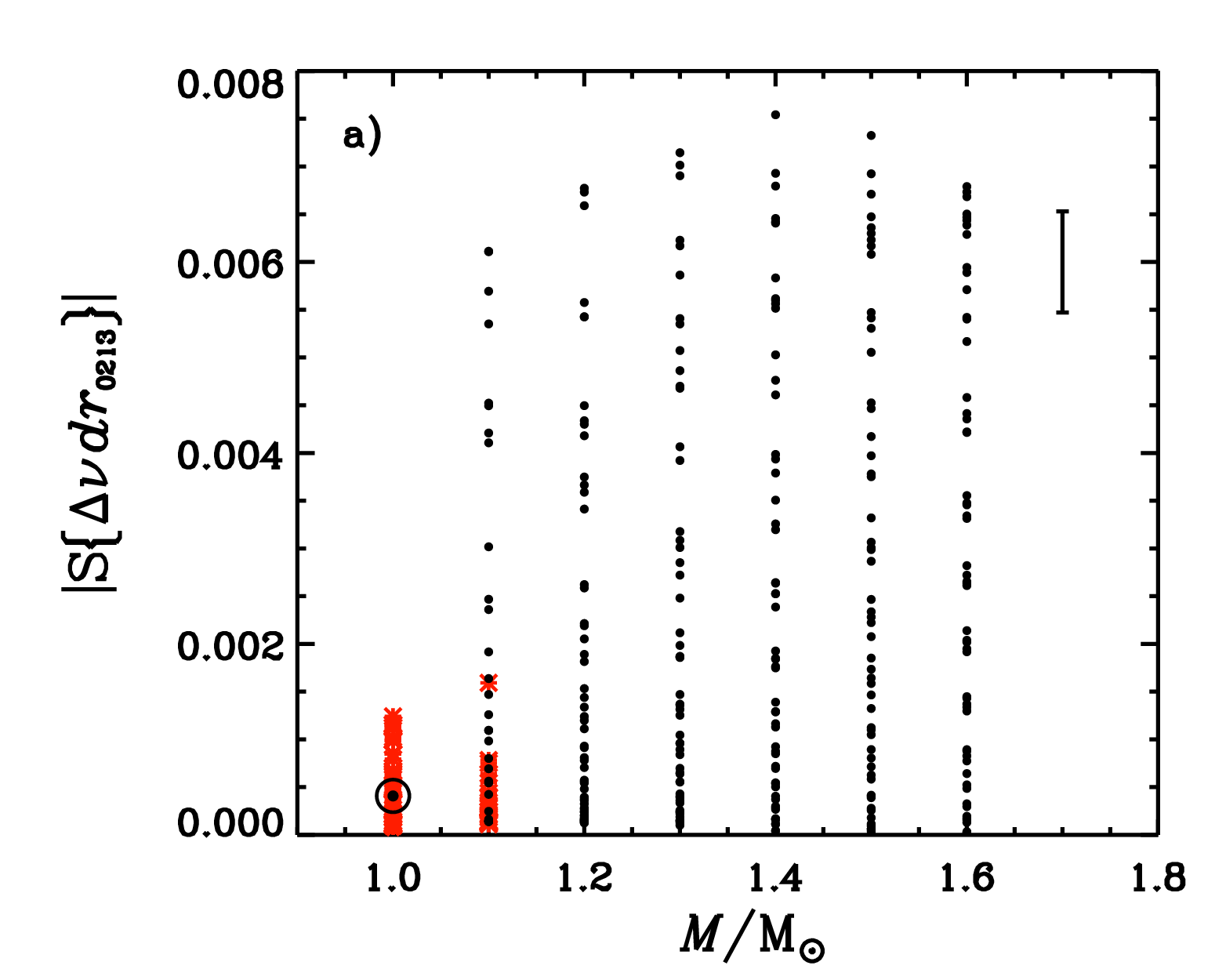} 
\includegraphics[width=9cm,angle=0]{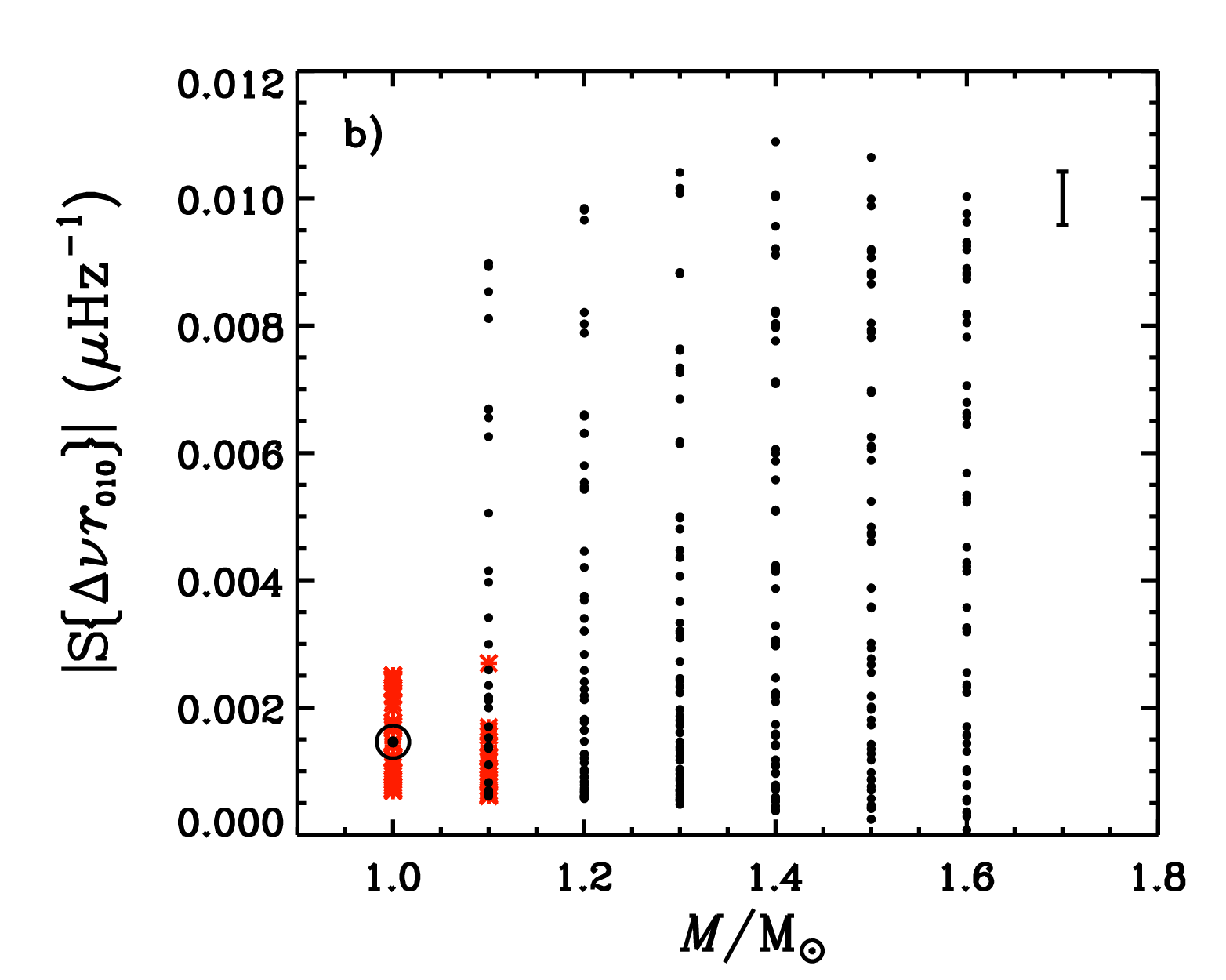} \\
\includegraphics[width=9cm,angle=0]{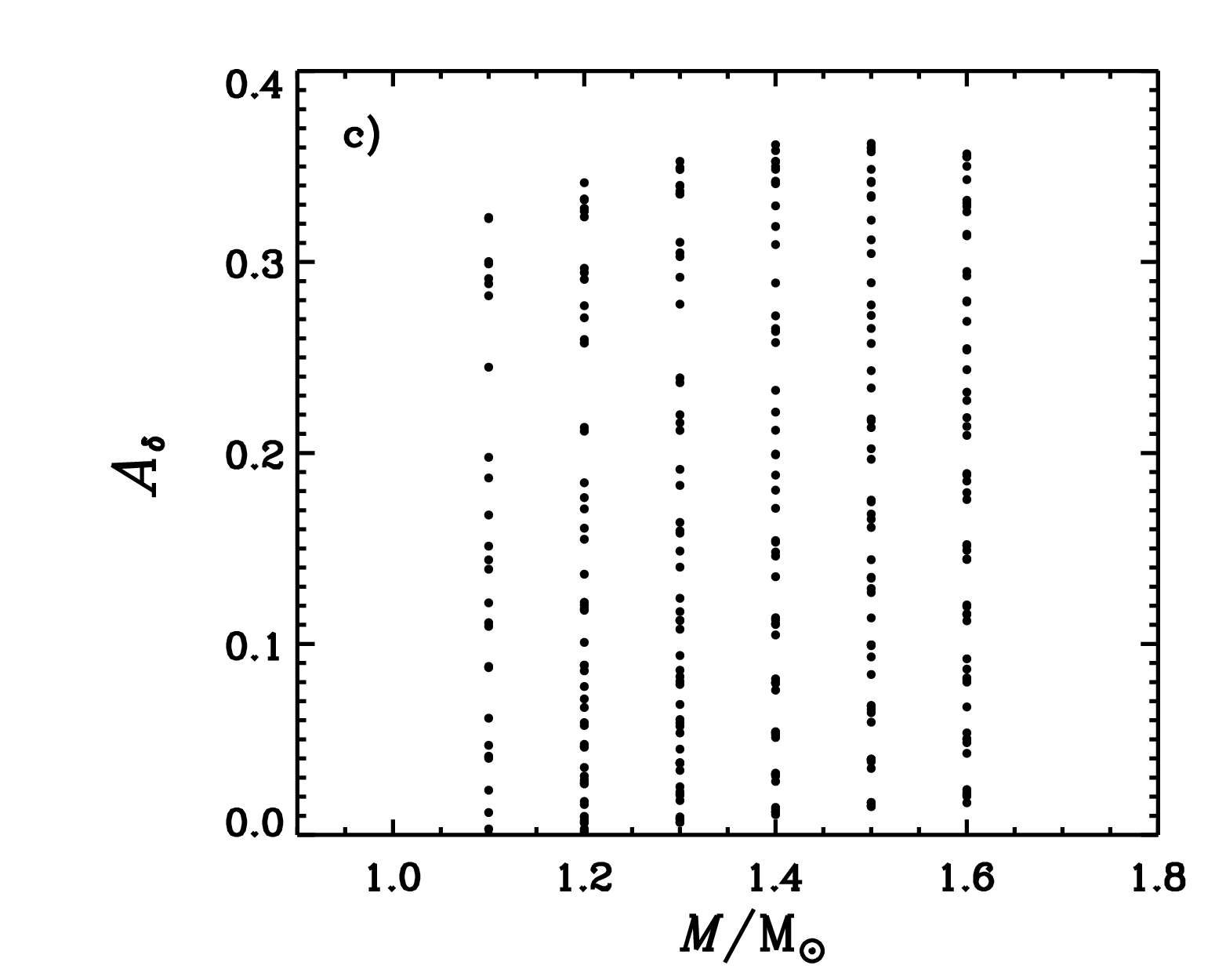}
\includegraphics[width=9cm,angle=0]{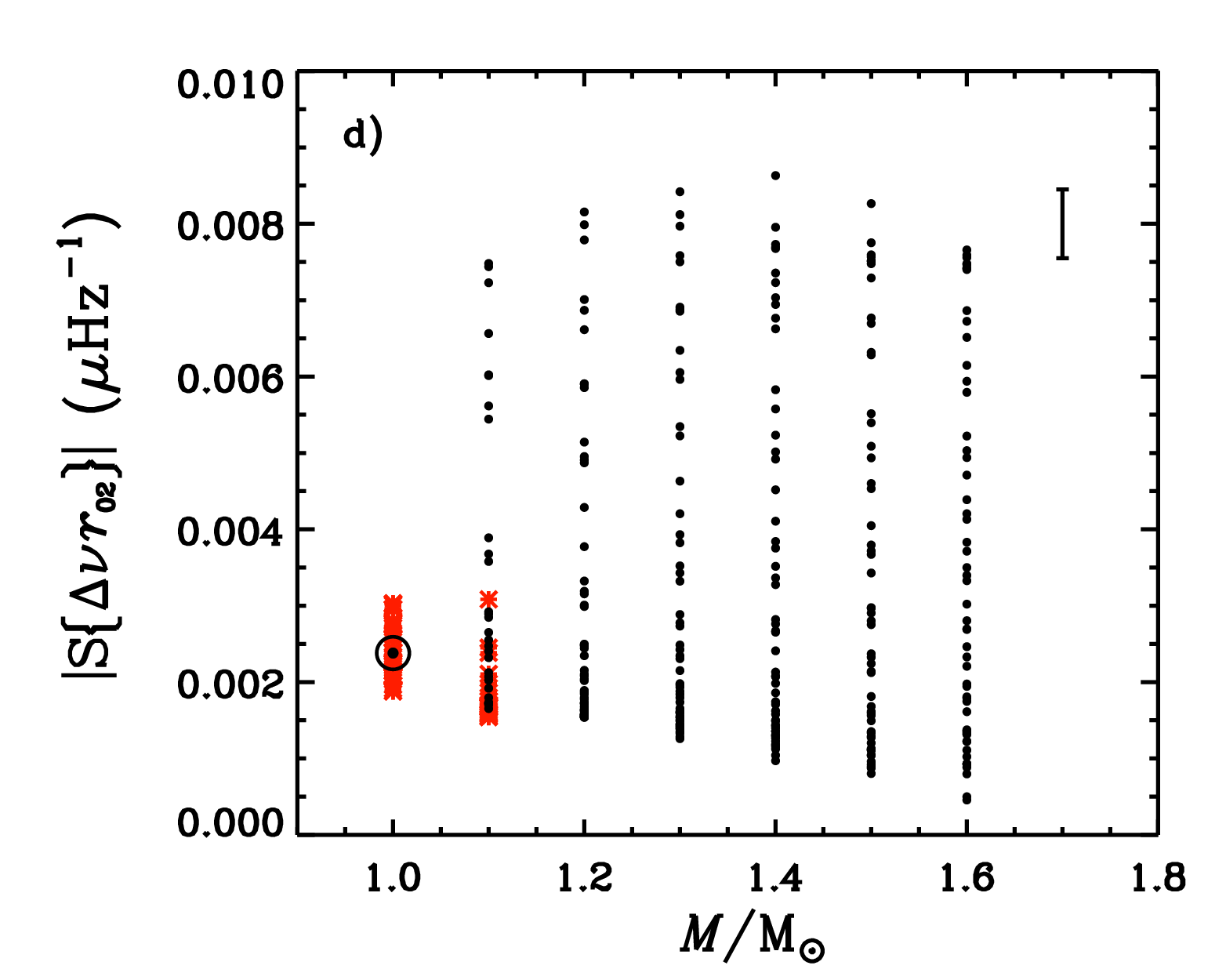} 
\end{minipage}
\caption{Panel (a): the symbols represent the absolute value of the slopes for the quantity $\Delta\nu\ dr_{0213}$,
 $\left| S\{\Delta\nu\,dr_{0213}\}\right|$,
for all models considered as a function of the mass of the model. The red stars correspond to models with no convective core.
The typical error bar for the slope of $\Delta\nu\,dr_{0213}$ is also shown in this plot.
Panel (c) shows, for the subset of models shown in panel (a) 
that have convective cores, the amplitude of the discontinuity as a function of the mass of the model.
Panels (b) and (d) are similar to panel (a), but for the quantities $\Delta\nu\,r_{010}$ 
and $\Delta\nu\,r_{02}$, respectively. The symbol of the Sun shown at $M = 1.0\,\rm M_{\sun}$
represents the slopes of the diagnostic tools computed using the solar frequencies from Model S \citep{cd96}.}
\label{fig:max_mass}
\end{figure*}
\subsection{Slopes versus discontinuity in sound speed}
\label{sec4.2}
According to \cite{cunha11}, the quantity $dr_{0213}$ is related to the 
frequency perturbation $\delta\nu^{\rm c}$ induced by the discontinuity in sound speed at 
the edge of the convective core by
\begin{equation}
\label{eq_deltanuc1}
 \delta\nu^{\rm c}  \sim  \Delta\nu_{n-1,1}\,dr_{0213}.
\end{equation}
Moreover, from the author's analysis it results that
\begin{equation}
\label{eq_deltanuc2}
\delta\nu^{\rm c} \sim - F\,A_\delta,
\end{equation}
where $F$ is a function that depends on frequency, 
as well as on the shape of the sound speed around the discontinuity.

\begin{figure*}
\begin{center}
$\begin{array}{c}
\includegraphics[width=6cm,height=5cm,angle=0]{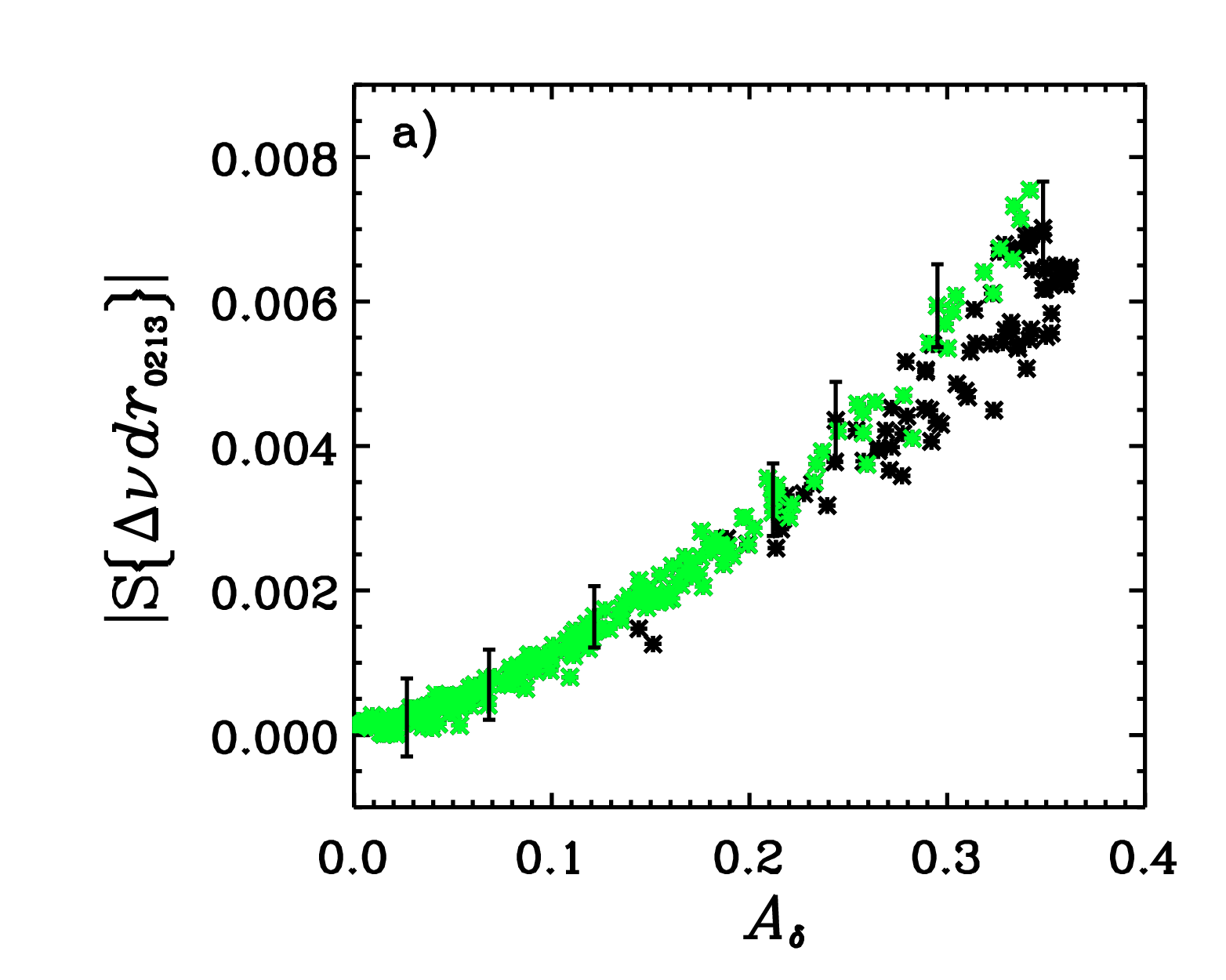} 
\includegraphics[width=6cm,height=5cm,angle=0]{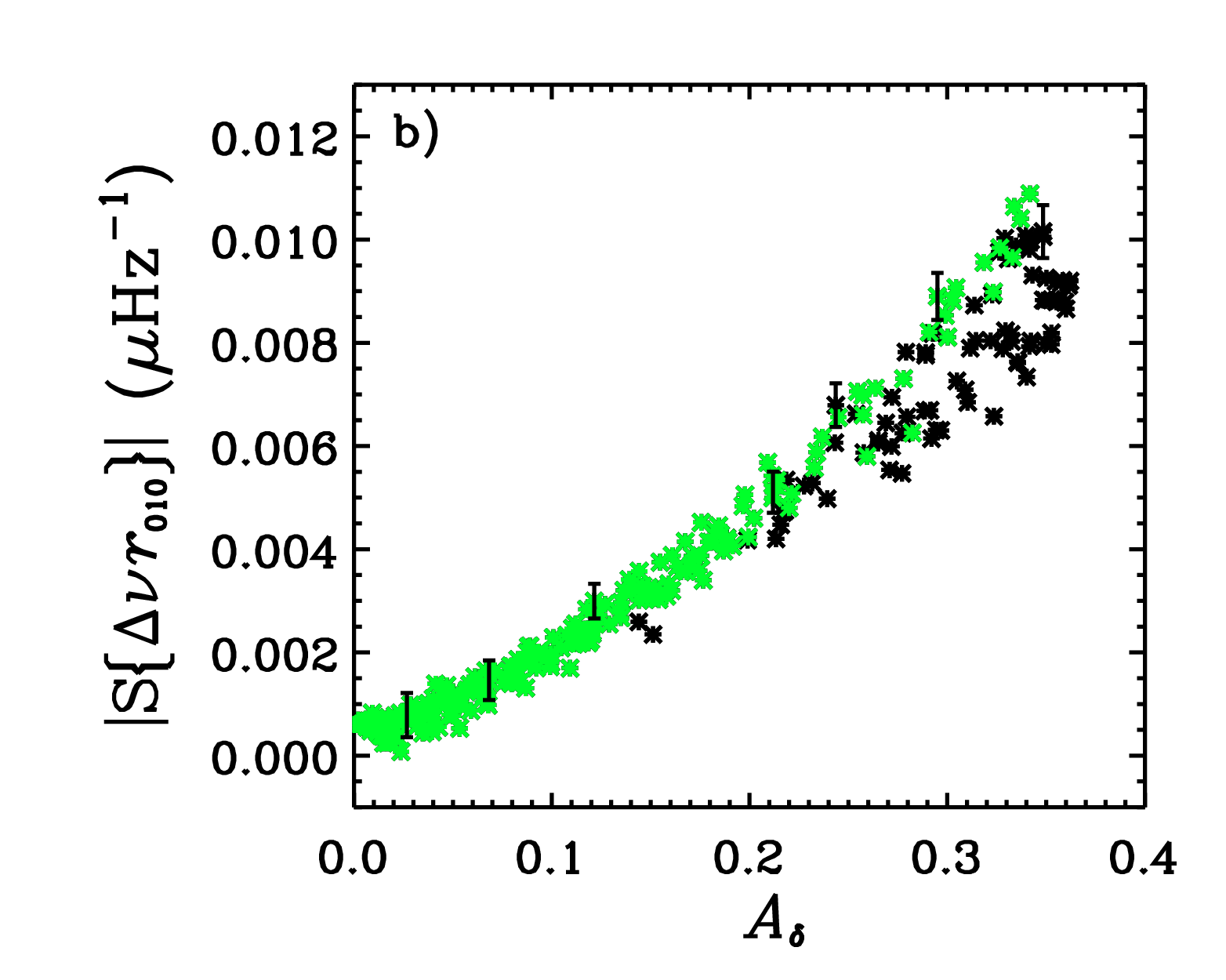} 
\includegraphics[width=6cm,height=5cm,angle=0]{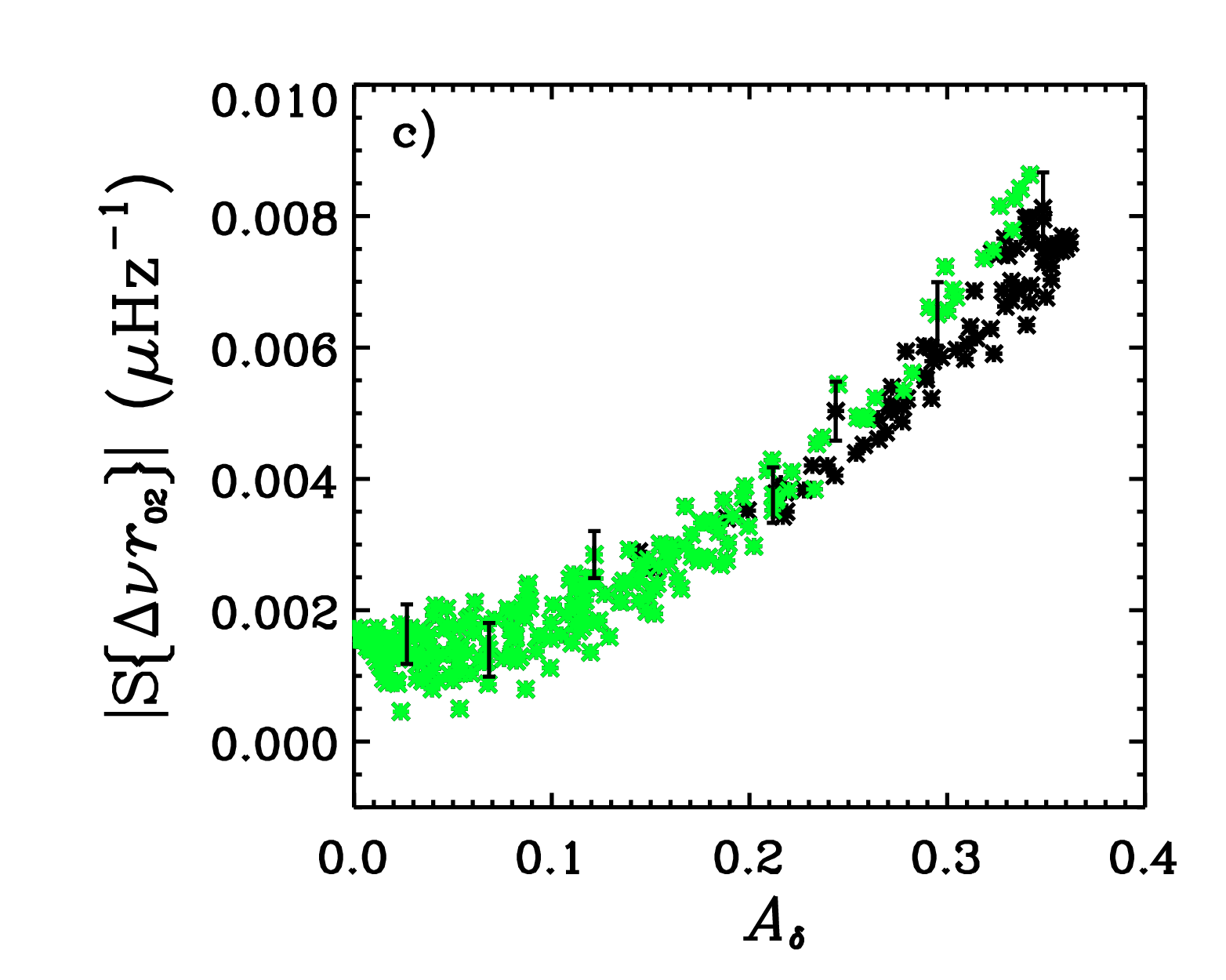} 
\end{array}$
\end{center}
\caption{The absolute value of the slopes of the diagnostic 
tools $\Delta\nu\,dr_{0213}$ (panel a), $\Delta\nu\,r_{010}$ (panel b)
and $\Delta\nu\,r_{02}$ (panel c) as a function of the relative amplitude $A_\delta$ of
the squared sound speed. Each point represented by a star symbol, 
either black or green, corresponds to one model of our grids.
Only models with a convective core are shown.
These include models with masses between 1.1 and 1.6$\,\rm M_{\sun}$  (see Section\,\ref{sec4.1} for details). 
The green stars represent the subset of models for which the slope of the diagnostic tools was measured between 
$\nu_{\rm max}$ and $\nu_{\rm c}$ (see Section\,\ref{sec4.3} for details).
The error bars shown were computed for some
of the models of our grid.}
\label{fig:slope_deltacc2}
\end{figure*}
\begin{figure*}
\begin{center}
$\begin{array}{c}
\includegraphics[width=6cm,height=5cm,angle=0]{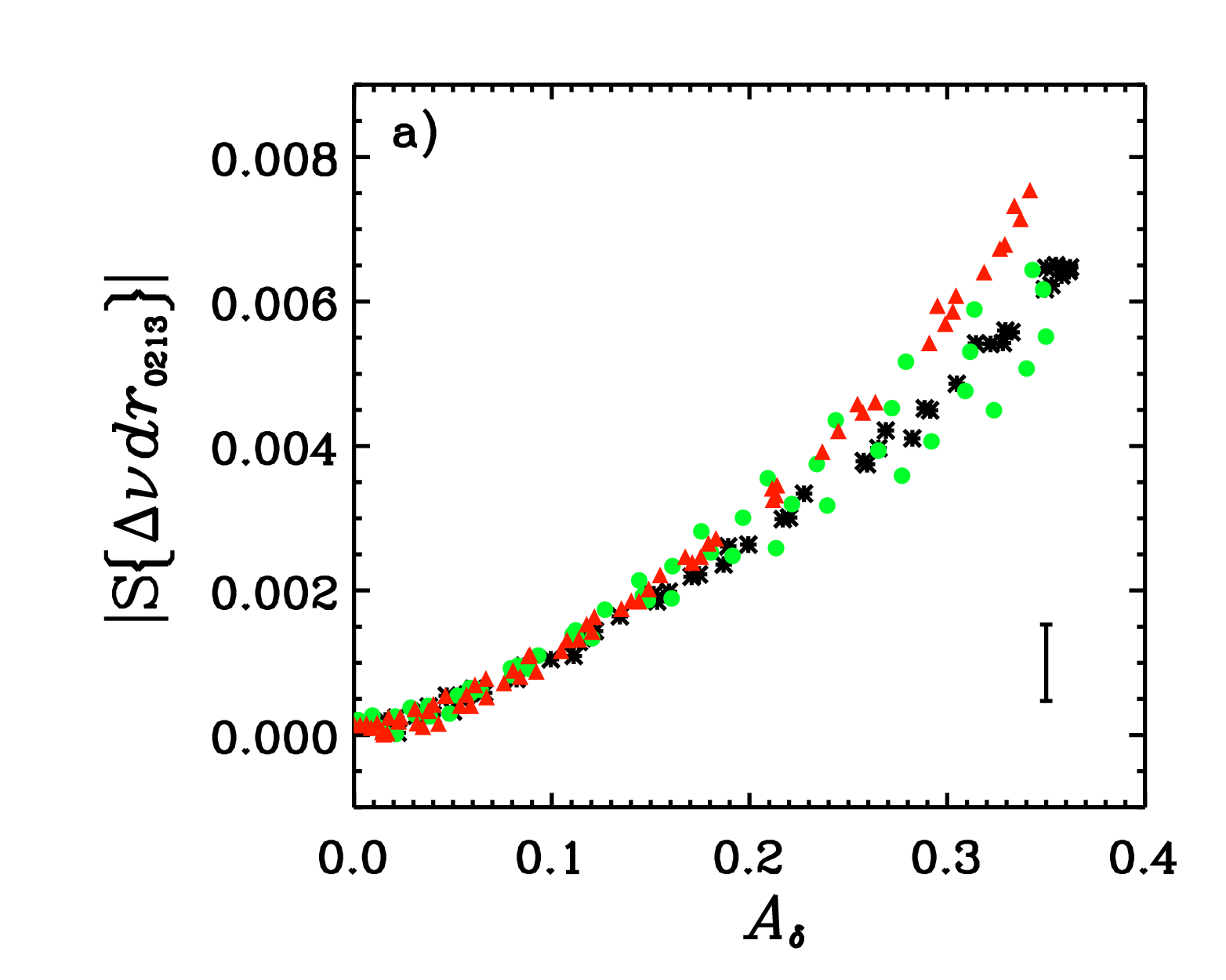}
 \includegraphics[width=6cm,height=5cm,angle=0]{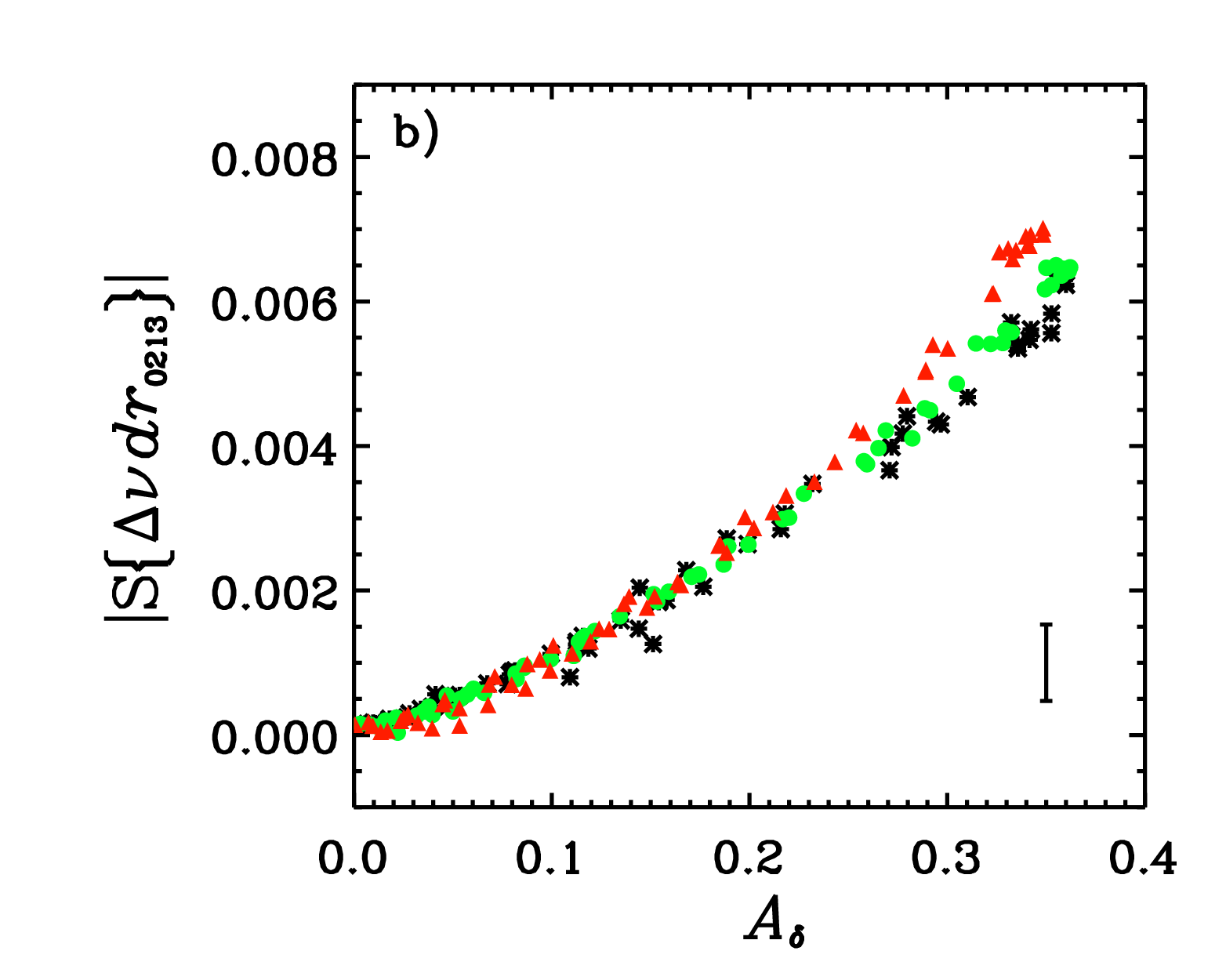}
\includegraphics[width=6cm,height=5cm,angle=0]{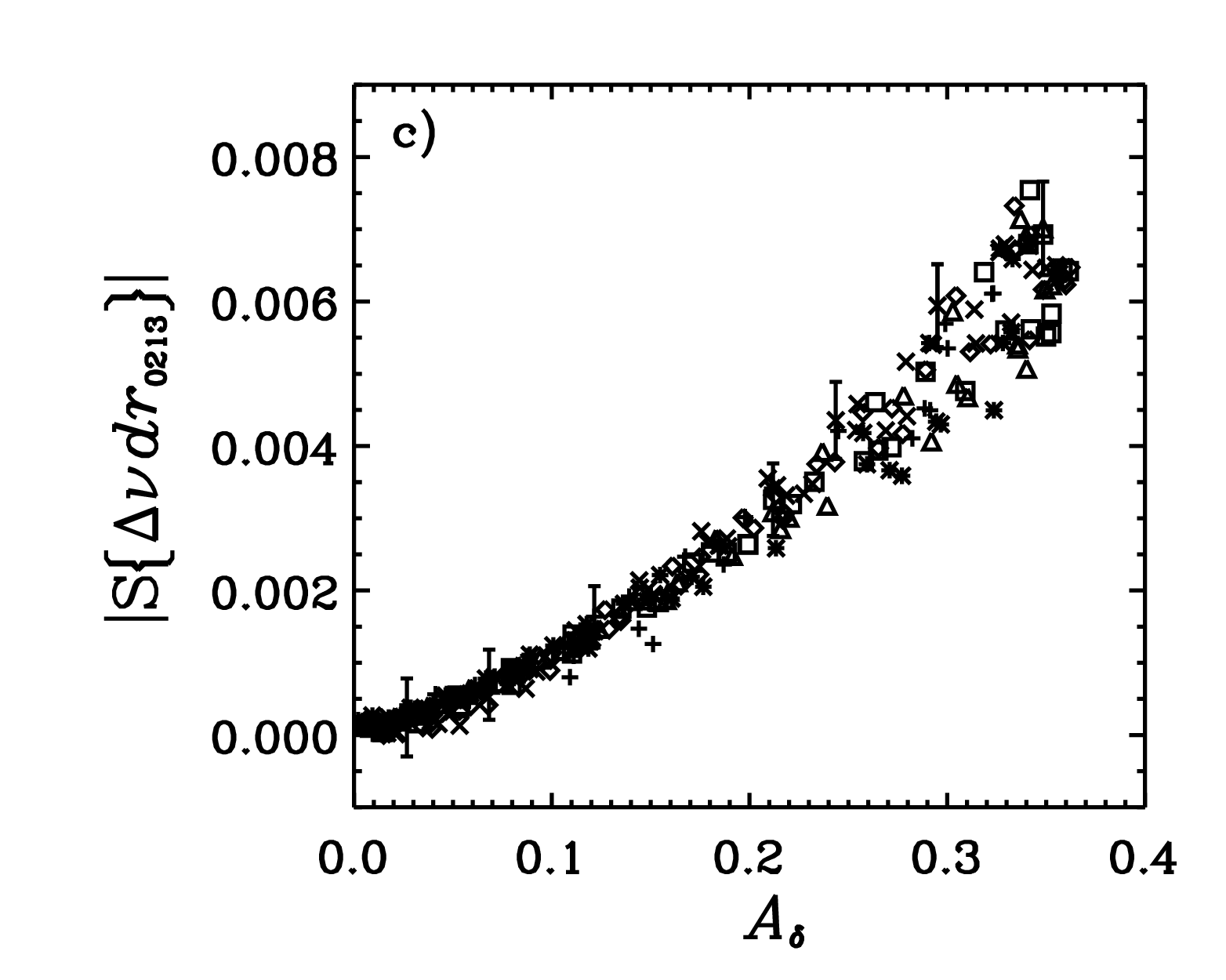}
\end{array}$
\end{center}
\caption{Panel (a): the absolute value of the slopes of the diagnostic tool $\Delta\nu\,dr_{0213}$ as a function
of the relative amplitude $A_\delta$ of the discontinuity in the squared sound speed.
The black stars represent models with solar metallicity, $Z/X=0.0245$,
the green filled circles represent models with $Z/X=0.0079$, and the red
triangles represent models with $Z/X=0.0787$. The models have $\alpha_{\rm OV}=0.1$.
The typical error bar for  $\left| S\{\Delta\nu\,dr_{0213}\}\right|$, is shown in the lower right corner of the plot.
Panel (b) shows the same quantities as in panel (a),
but here the black stars represent models without overshoot, $\alpha_{\rm OV}=0.0$,
the green filled circles represent models with $\alpha_{\rm OV}=0.1$ and the red
triangles represent models with $\alpha_{\rm OV}=0.2$. The models have $Z/X=0.0245$.
Panel (c) also shows the same quantities as in panel (a) but here the different symbols correspond to different values of the mass, namely
$M = 1.1\,\rm M_{\sun}$ are represented by crosses, $M = 1.2\,\rm M_{\sun}$ by stars,
$M = 1.3\,\rm M_{\sun}$ by triangles, $M = 1.4\,\rm M_{\sun}$ by squares,
$M = 1.5\,\rm M_{\sun}$ by diamonds and $M=1.6\,\rm M_{\sun}$ by X. The error
bars for  $\left| S\{\Delta\nu\,dr_{0213}\}\right|$ in this case are shown for some models.}
\label{fig:slope_deltacc2_met}
\end{figure*}
Let's now consider the quantity $ S\{\Delta\nu\,dr_{0213}\}$ previously defined.
Then, we have,
\begin{equation}
 S\{\Delta\nu\,dr_{0213}\} \approx \left[ \frac{\textrm{d}}{\textrm{d}\nu}\left(\Delta\nu\,dr_{0213}\right)\right]_{\nu_{\rm slope}}, 
\end{equation} 
where the derivative of the diagnostic tool $\Delta\nu\,dr_{0213}$ is to be taken at a place 
where $F$ varies linearly with $\nu$ (which is approximately true at the 
frequency $\nu_{\rm slope}$ where the slope is computed) and where $\Delta\nu$ is the average 
large separation for modes of degree $l = 1$ in the frequency range where the derivative is to be taken. 
From relations (\ref{eq_deltanuc1}) and (\ref{eq_deltanuc2}), one thus expects
that $\left| S\{\Delta\nu\,dr_{0213}\}\right|$ provides a measure of the
sound-speed discontinuity at the edge of any core.  

In Fig. \ref{fig:max_mass}(a), we show
$\left| S\{\Delta\nu\,dr_{0213}\}\right|$
computed for all models within our sequence
of evolutionary tracks, as a function of the mass of the models. 
Models without a convective core are shown by the asterisk symbol in red.
Fig. \ref{fig:max_mass}(c)
shows the amplitude $A_\delta$ of the discontinuity in the squared sound speed
as a function of the mass of the models. Note that no stars of 
$M = 1.0\,\rm M_{\sun}$ are shown in this plot,
because these stars do not have convective cores, and hence have
no discontinuity in the sound speed.
From panel (a) of this figure, we can see that 
for the physics considered in the models,
$\left| S\{\Delta\nu\,dr_{0213}\}\right|$
is no higher than $\sim$0.008, independently of the mass of the models.
This maximum value of $\sim$0.008 is associated with
a maximum value for $A_\delta$ of $\sim$0.4,
which occurs when stars approach the TAMS and have a core 
essentially composed of helium.  
The existence of these maxima results
from the fact that the variation of 
the mean molecular weight at the edge of the core is itself limited. 

In Fig. \ref{fig:max_mass}(a),
we can also see that models with $\left| S\{\Delta\nu\,dr_{0213}\}\right| \gtrsim 0.002$
all have a convective core, while models with 
$\left| S\{\Delta\nu\,dr_{0213}\}\right| \lesssim 0.002$
may or may not have a convective core, depending on the mass. Therefore, 
for a given observation of a star,
if we find that $\left| S\{\Delta\nu\,dr_{0213}\}\right| \gtrsim 0.002$,
we can say with confidence that the star has a convective core.

If the $l=3$ modes are not observed,
the two diagnostic tools $d_{010}$ and $d_{02}$, or their respective ratios,
which consider modes of degree only up to 2 should be preferred. 
Note, however, that these two quantities measure differently 
the structure of the core; hence, unlike $dr_{0213}$, they do not isolate
the signature of the sharp structural variation in the sound speed. 
As a consequence, when using these two diagnostic
tools one should have in mind that the effect of 
a much wider region of the star, and not only the discontinuity, is present. 
Nevertheless, it is interesting to see that
when analysing the slopes of 
$\Delta\nu\,r_{02}$ and $\Delta\nu\,r_{010}$ 
we obtain results that are analogous to the ones previously
mentioned for the diagnostic tool $\Delta\nu\ dr_{0213}$. 
We verified that $\left| S\{\Delta\nu\,r_{010}\}\right|$ and
$\left| S\{\Delta\nu\,r_{02}\}\right|$ are also bounded, 
in this case by maximum values  of 0.011 and $0.009\,\mu \rm Hz^{-1}$, respectively 
(cf. Fig. \ref{fig:max_mass}\,panels b and d).
Moreover, models with $\left| S\{\Delta\nu\,r_{010}\}\right|\gtrsim0.003\,\mu \rm Hz^{-1}$
or  $\left| S\{\Delta\nu\,r_{02}\}\right|\gtrsim0.0035\,\mu \rm Hz^{-1}$ all have convective cores.
So, as before, for these two quantities, we are confident that 
a detection of a slope with
absolute value larger than $\sim$0.0035$\,\mu \rm Hz^{-1}$
is a strong indication of the presence 
of a convective core. 
\begin{figure*}
\centering
\begin{minipage}{\linewidth}
$\begin{array}{c}
\includegraphics[width=9cm,angle=0]{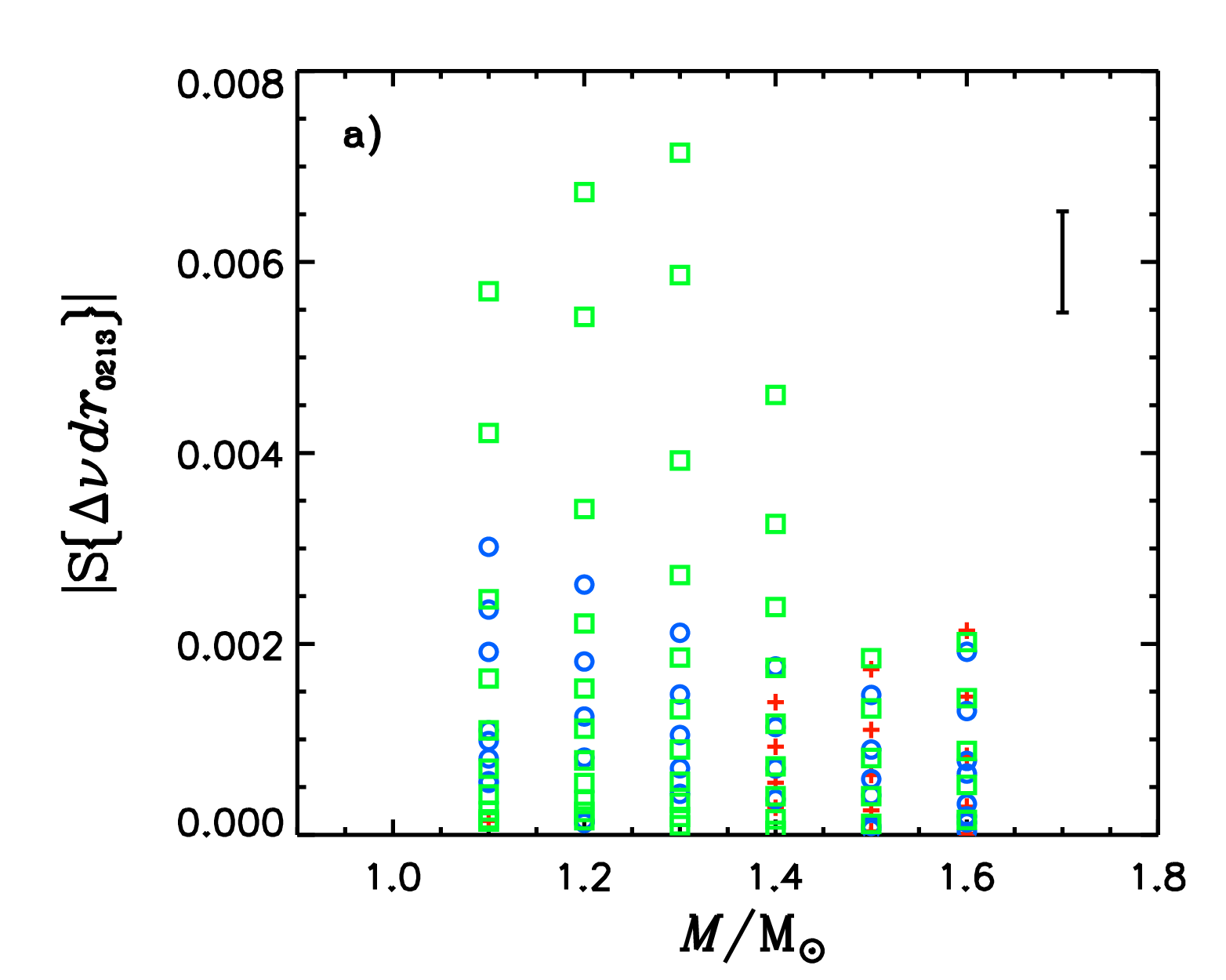}
\includegraphics[width=9cm,angle=0]{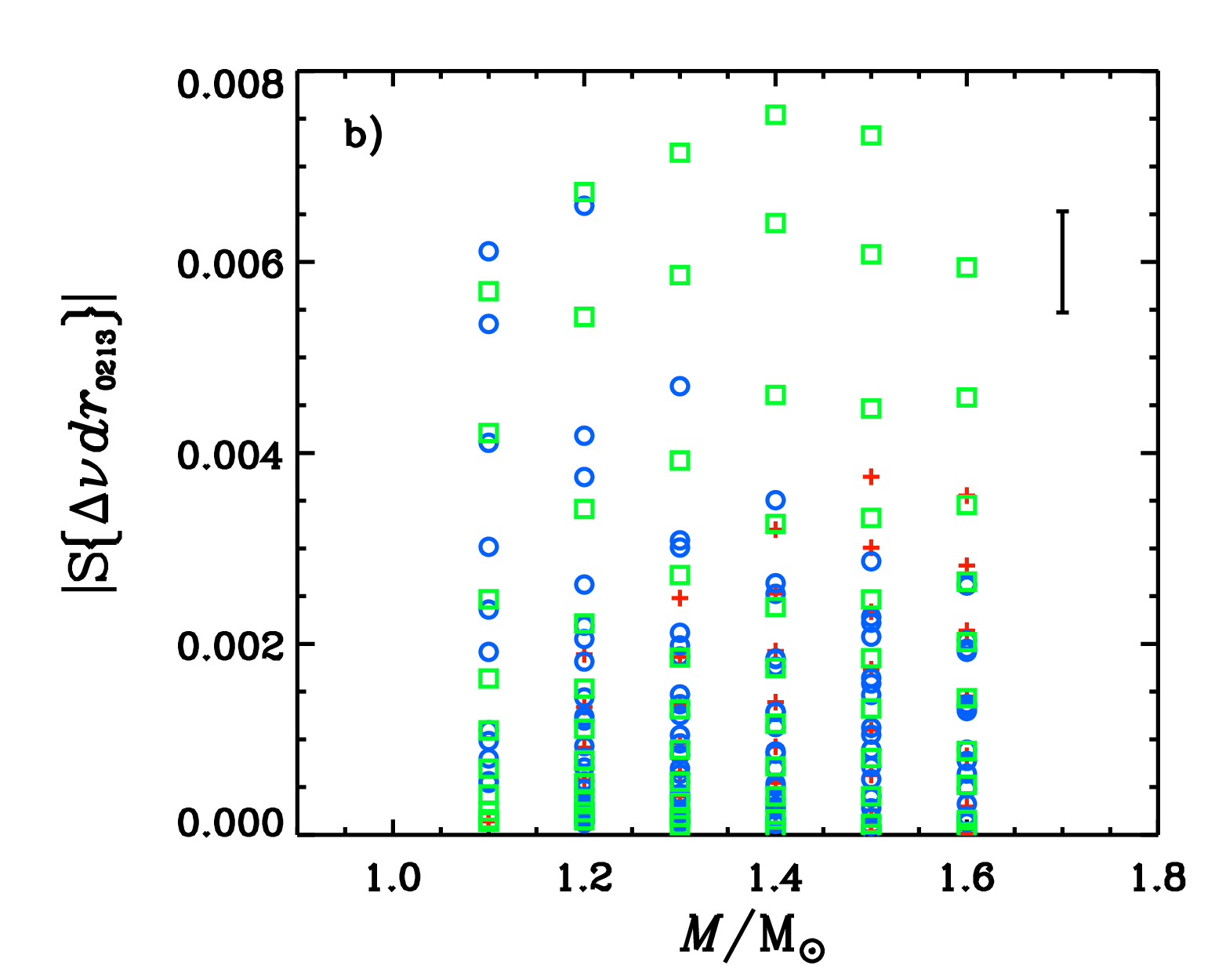}
\end{array}$
\end{minipage}
\caption{The same as in Fig. \ref{fig:max_mass}(a) but considering 
only models for which the frequency region where the slopes are computed is 
between $\nu_{\rm max}-8\Delta\nu$ and $\nu_{\rm max}+8\Delta\nu$ (panel a) or between
$\nu_{\rm max}$ and $\nu_{\rm c}$ (panel b). The different symbols represent
different metallicities, namely red crosses represent $Z/X=0.0079$, blue circles represent 
$Z/X=0.0245$ and green squares represent $Z/X=0.0787$.}
\label{fig:max_mass2}
\end{figure*}

Next, we investigate in more detail the relation between 
the slopes and the amplitude of the discontinuity. 
If the function $F$ were model independent, then one would expect
$\left| S\{\Delta\nu\,dr_{0213}\}\right| \propto A_\delta$,
i.e., one would expect a linear relation between
$\left| S\{\Delta\nu\,dr_{0213}\}\right|$ and $A_\delta$.
However, the fact that $F$ is model-dependent means that 
deviations from a simple linear relation, as well as significant dispersion, 
may exist in the relation between these two quantities.  

In Fig. \ref{fig:slope_deltacc2}(a),
we show the relation between the absolute value of the slopes computed 
for the diagnostic tool $\Delta\nu\,dr_{0213}$, 
$\left| S\{\Delta\nu\,dr_{0213}\}\right|$, and
the relative amplitude $A_\delta$ of the discontinuity in the squared sound speed.
As anticipated from the work of \cite{cunha11}, 
there is  a strong dependence of the slopes on
the amplitude of the discontinuity. Nevertheless, that 
relation deviates from linear and shows significant dispersion, 
particularly for larger values of $A_\delta$. 
This may be explained by the fact that as the star approaches 
the TAMS, the sound-speed perturbation becomes wider 
(cf. Fig. \ref{fig:c2_rr_m}, bottom panel) 
and the model dependence of the function $F$ becomes more evident.  

In fact, for the younger stars with $A_\delta\lesssim0.2$, the relation
shown in panel a) of Fig. \ref{fig:slope_deltacc2} was found not to depend significantly on the mass, core overshoot
or metallicity, at least for the physics that we considered in our set of models. 
However, at latter stages, that is no longer the case. To illustrate this, we show,
in Fig. \ref{fig:slope_deltacc2_met}(a), the dependence
of the relation between $\left| S\{\Delta\nu\,dr_{0213}\}\right|$
and $A_\delta$ on the metallicity. In Fig. \ref{fig:slope_deltacc2_met}(b)
we show the dependence of the same relation on the overshoot and, finally, 
in Fig. \ref{fig:slope_deltacc2_met}(c) we show the dependence on the mass.
A dependence of the relation on metallicity and overshoot clearly emerges
as the stars approach the TAMS. On the other hand, no clear dependence
on stellar mass is seen even at the latest evolution stages.  

In Figs.\,\ref{fig:slope_deltacc2}(b) and (c),
we show the same relation as before, but for the quantities 
$\Delta\nu\,r_{010}$ and $\Delta\nu\,r_{02}$, respectively. 
Although the two latter diagnostic tools
do not isolate the effect of the discontinuity in the sound speed, the similarity
with the plot for the first diagnostic tool, particularly for the case of $\Delta\nu\,r_{010}$, 
indicates that their slopes are strongly affected by this discontinuity.
\subsection{Diagnostic potential}
\label{sec4.3}
To establish the relations discussed in Section \ref{sec4.2}, 
between the absolute value of the slopes of the diagnostic tools and the amplitude of the discontinuity 
in the squared sound speed, we have considered all models in our grid. 
However, as mentioned earlier, for a number of models, the frequency 
at which the slope is computed is not in the range of 
observed frequencies. For instance, for a typical main-sequence 
solar-like pulsator observed by the \textit{Kepler} satellite, only about 
a dozen of radial orders, centred on  $\nu_{\rm max}$,  are observed. 

To illustrate the impact of these observational limitations, 
we have considered two subset of models, namely 
(1) a subset composed of models for which the frequency region 
where the slopes are computed is between $\nu_{\rm max}-8\Delta\nu$ and $\nu_{\rm max}+8\Delta\nu$ 
and (2) a subset composed of models for which the frequency 
region where the slopes are computed is between $\nu_{\rm max}$ and $\nu_{\rm c}$.
They contain, respectively, 35 and 74 per cent of the total number of stars 
with a convective core in our sample. These percentages increase
to 49 and 98 per cent of the models in our sample, respectively, when we consider only models 
with a convective core and $A_\delta \le 0.2$.

We show in Fig. \ref{fig:max_mass2} again the
plot present in Fig. \ref{fig:max_mass}(a) but considering only models
for which the frequency region where the slopes are computed is 
between $\nu_{\rm max}-8\Delta\nu$ and $\nu_{\rm max}+8\Delta\nu$ (panel a) or between
$\nu_{\rm max}$ and $\nu_{\rm c}$ (panel b). The different
metallicities considered in the models are shown with different symbols.
Models with  $\left| S\{\Delta\nu\,dr_{0213}\}\right| \gtrsim 0.003$
and for which the frequency region where the slopes are computed is 
between $\nu_{\rm max}-8\Delta\nu$ and $\nu_{\rm max}+8\Delta\nu$
all have the highest metallicity, namely $Z/X=0.0787$. 

In Fig. \ref{fig:slope_deltacc2}, the green symbols represent
the models for which the slope of the diagnostic tools
was measured between $\nu_{\rm max}$ and $\nu_{\rm c}$.
Clearly, when considering only the subset of models for which 
the frequency region where the slopes are computed may be observed, the dispersion in the 
relation between the slopes of the diagnostic tools and the amplitude of the discontinuity 
in the sound speed squared seen earlier for the more evolved models is significantly reduced.
Thus, from our results we may conclude that it is
possible, for a significant subset of stars, to get a measure of
the amplitude of the discontinuity in the sound speed
from the analysis of the observed oscillation frequencies.
Nevertheless, we note that this subset is either strongly 
biased towards stars in first half of their main-sequence lifetime
or towards stars with large metallicity. 

\begin{figure}
 \begin{center}
 \includegraphics[width=9cm,angle=0]{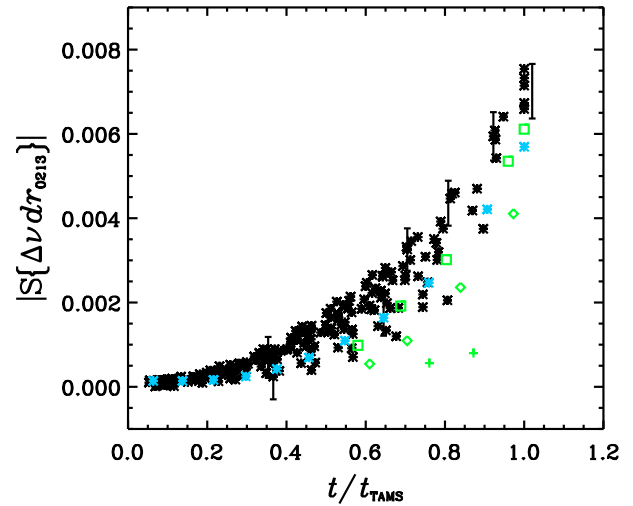}
 \end{center}
 \caption{The absolute value of the slopes of the diagnostic tool $\Delta\nu\ dr_{0213}$,
$\left| S\{\Delta\nu\,dr_{0213}\}\right|$, as a function
of the fraction of stellar evolution, $t/t_{\rm TAMS}$, for the models of
our grid that have a convective core and for which the frequency
region where the slopes are computed is between $\nu_{\rm max}$ and $\nu_{\rm c}$. 
These are all models with $1.2 \leq M \leq 1.6 \rm\,M_{\sun}$
and models with $M = 1.1\,\rm M_{\sun}$ with metallicities $Z/X=0.0245$ (green symbols) 
and $Z/X=0.0787$ (light blue stars). The different symbols in green represent
different values for the overshoot parameter, namely $\alpha_{\rm OV}$ = 0.0 (crosses),
$\alpha_{\rm OV}$ = 0.1 (diamonds) and $\alpha_{\rm OV}$ = 0.2 (squares).}
 \label{fig:slopedr0213_tttams_all}
 \end{figure}

\begin{figure}
 \begin{center}
 \includegraphics[width=9cm,angle=0]{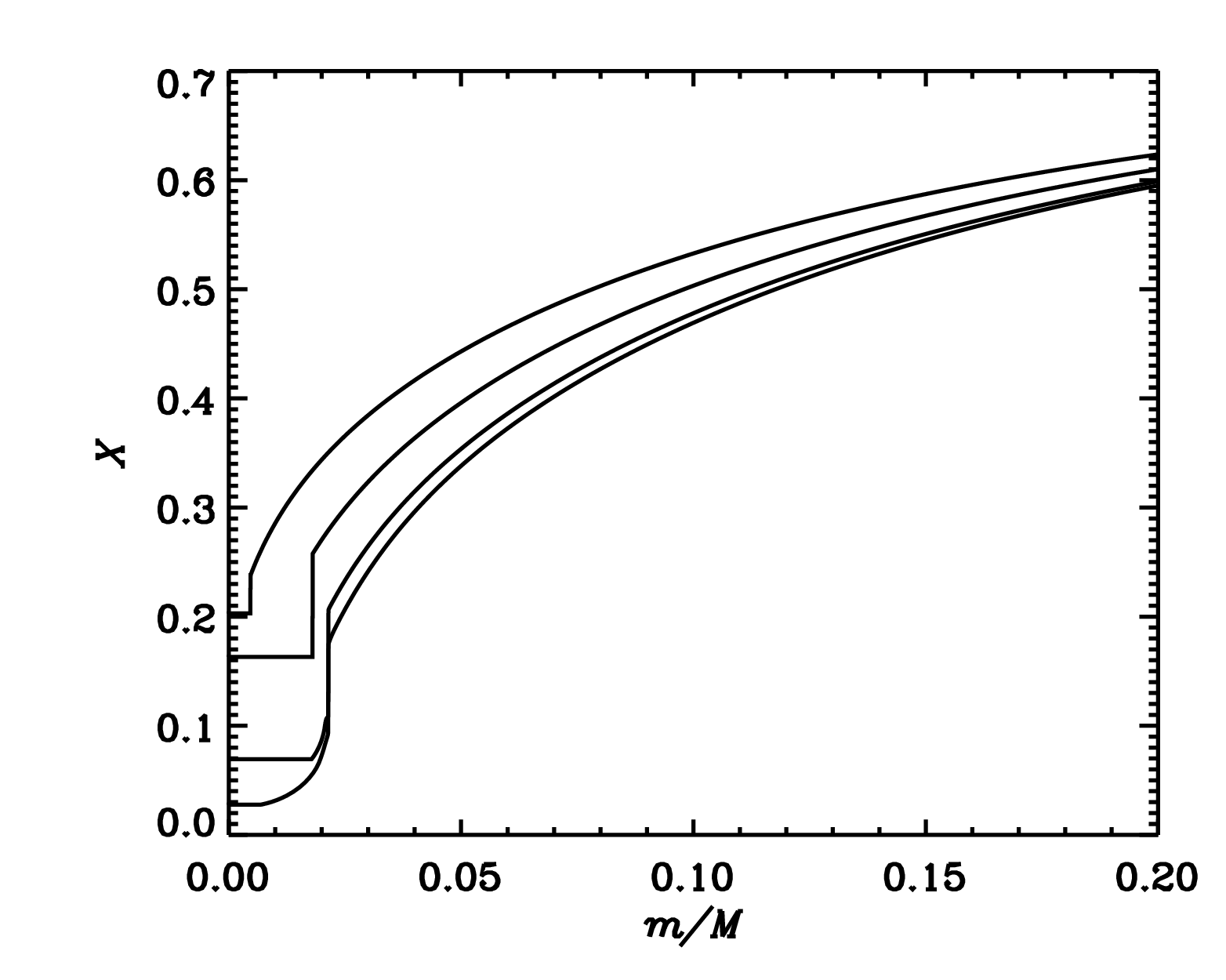}
 \includegraphics[width=9cm,angle=0]{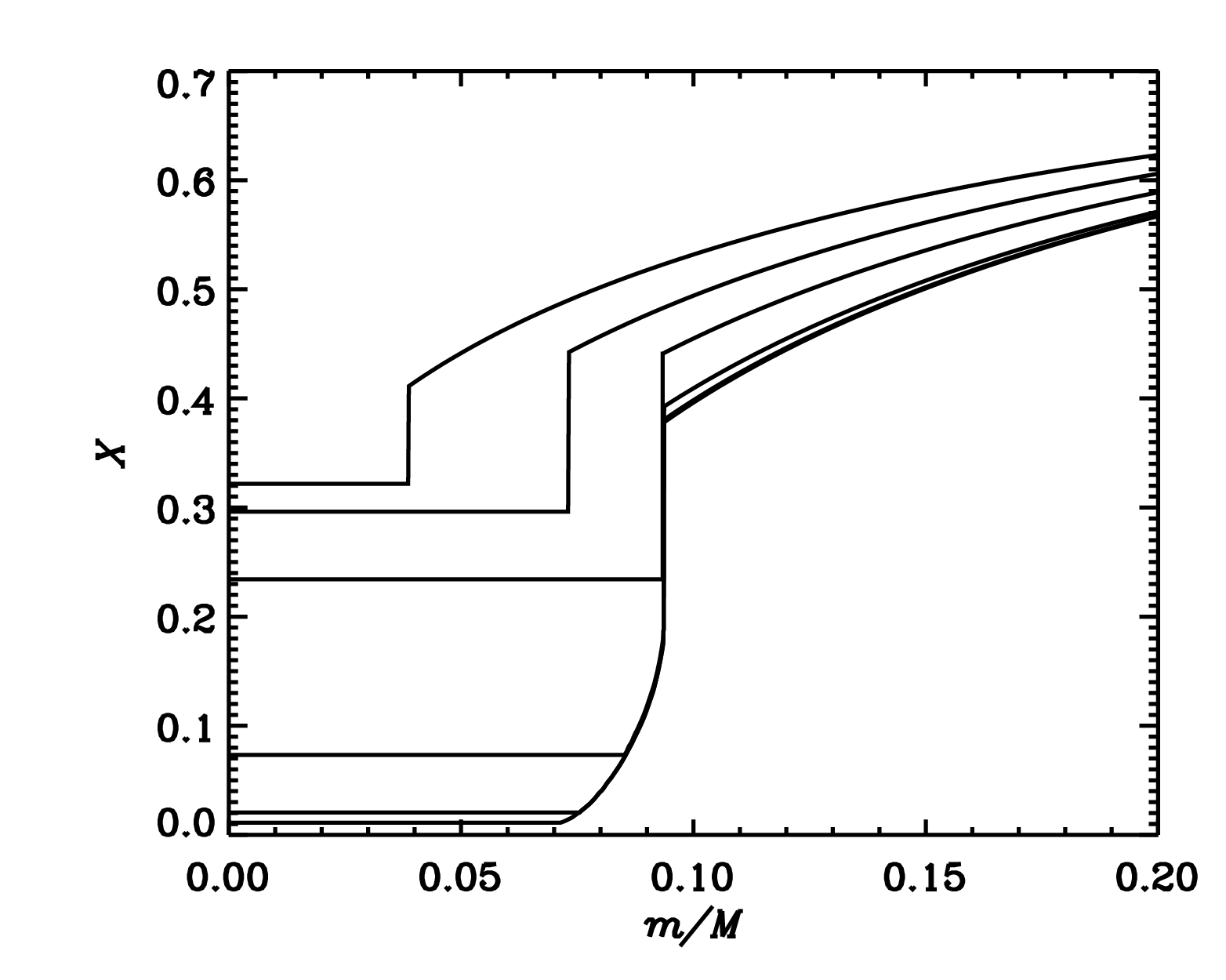}
 \end{center}
 \caption{The hydrogen profile of a $M=1.1\rm\,M_{\sun}$ model with $Z/X=0.0245$
and with $\alpha_{\rm OV}$ = 0.0 (upper panel) and a $M=1.1\rm\,M_{\sun}$ model with $Z/X=0.0245$
and with $\alpha_{\rm OV}$ = 0.2 (lower panel). The different curves correspond 
to different stages of evolution. Only models with a convective core are shown.}
 \label{fig:x_mm_ov0-2_11}
 \end{figure}

One would hope that the measurement 
of the discontinuity in the sound speed
could be related to the evolutionary state of the star.
To test this possibility, we inspected directly
the relation between the slopes of the different
diagnostic tools and the star's fraction of evolution ($t/t_{\rm TAMS}$)
along the main sequence, where $t$ is the age of the star at a given 
evolutionary stage in the main sequence and $t_{\rm TAMS}$ is the age of the star
within the same evolutionary track but at the TAMS. 
We considered the stellar fraction of evolution and not 
the stellar age alone since the latter is strongly dependent
on the stellar mass.
Fig. \ref{fig:slopedr0213_tttams_all} shows, for all
models with a convective core and for which the frequency
region where the slope is computed is between $\nu_{\rm max}$ and $\nu_{\rm c}$,
the absolute value of the slopes of $\Delta\nu\,dr_{0213}$, 
$\left| S\{\Delta\nu\,dr_{0213}\}\right|$, as a function 
of the fraction of main-sequence stellar evolution, $t/t_{\rm TAMS}$. 
By inspecting this figure, we see a large spread in the slopes 
at the higher evolution fractions, namely at $t/t_{\rm TAMS}\gtrsim0.6$.
This is due to the fact that models with $M = 1.1\,\rm M_{\sun}$,
solar metallicity and $\alpha_{\rm OV}=0.0$ have a very small central
convective region. As a result, substantial conversion 
of hydrogen into helium takes place 
outside the inner convective region and, in turn, the 
discontinuity in the chemical composition 
at the edge of that convective region is
significantly smaller than in otherwise similar
models with overshoot, at the latest stages of evolution. 
This is illustrated
in Fig. \ref{fig:x_mm_ov0-2_11} where we compare the cases of models
with $M = 1.1\,\rm M_{\sun}$ and solar metallicity and different values 
for the overshoot parameter.
Moreover, we find that, for a similar reason, 
models with $M = 1.2\,\rm M_{\sun}$ and no overshoot
still have a relatively small convective core, hence a slope still smaller
than that of similar mass models with overshoot or models of
higher mass at the later stages of their evolution.
\begin{figure}
\begin{center}
\includegraphics[width=9cm,angle=0]{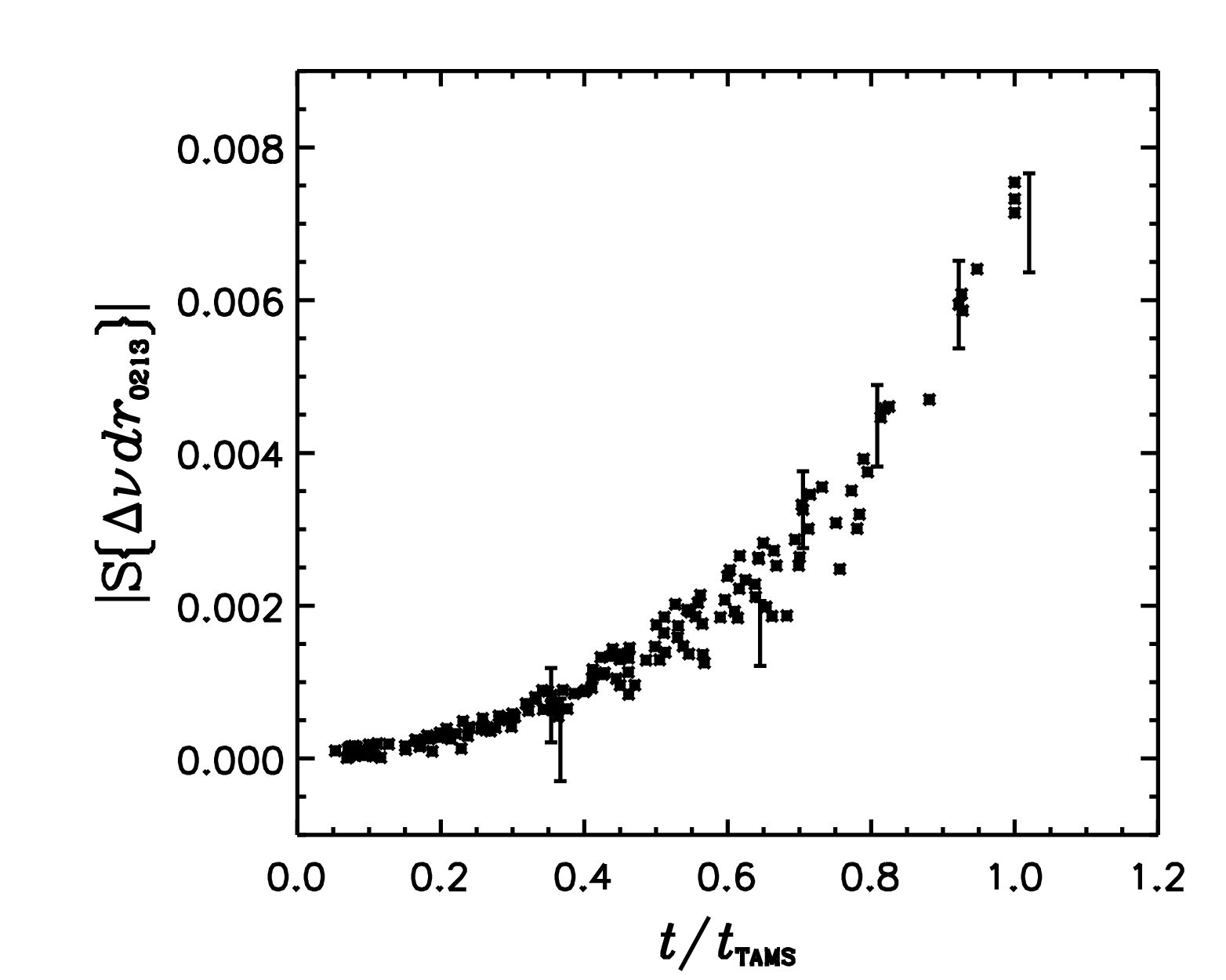} 
\includegraphics[width=9cm,angle=0]{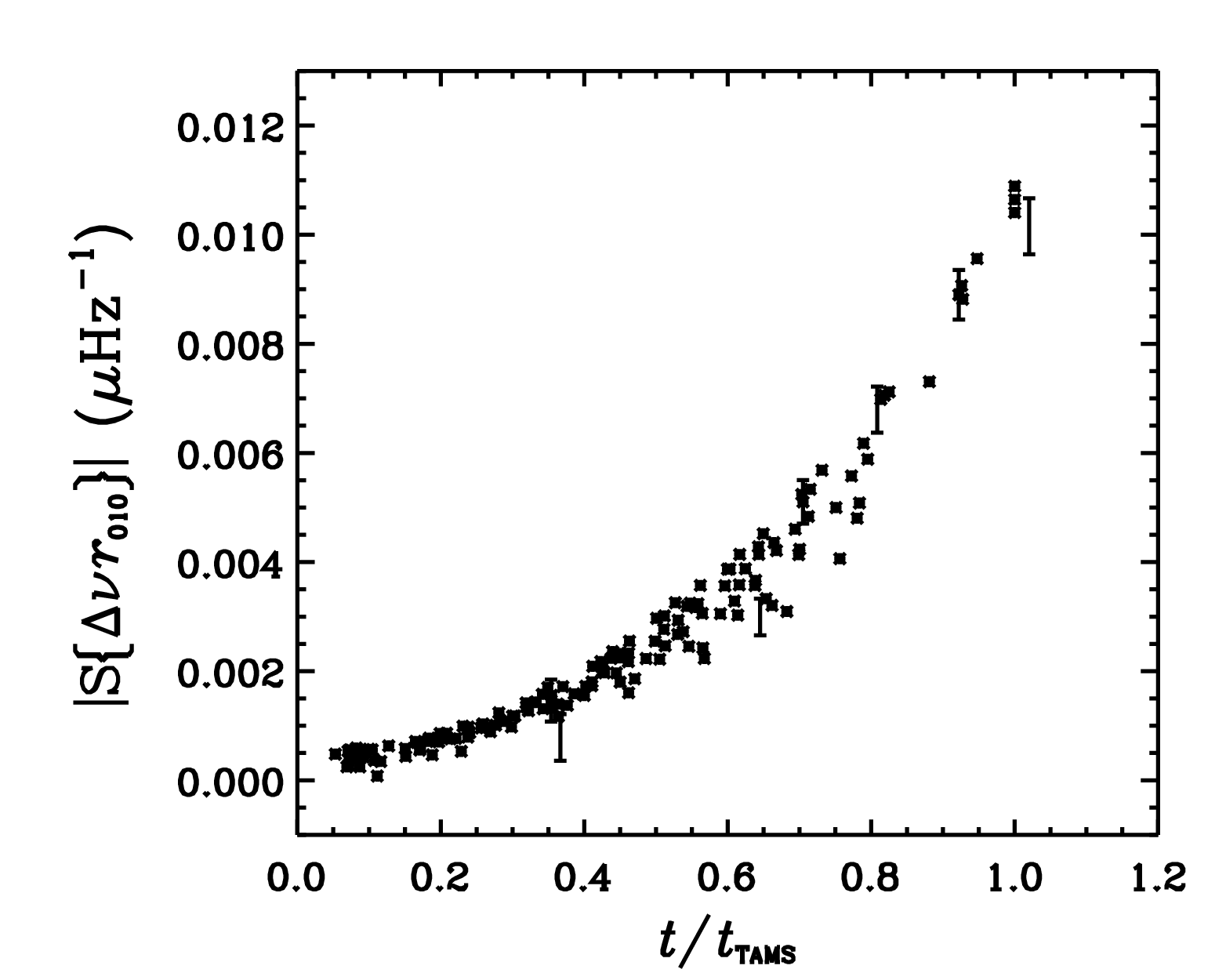} 
\includegraphics[width=9cm,angle=0]{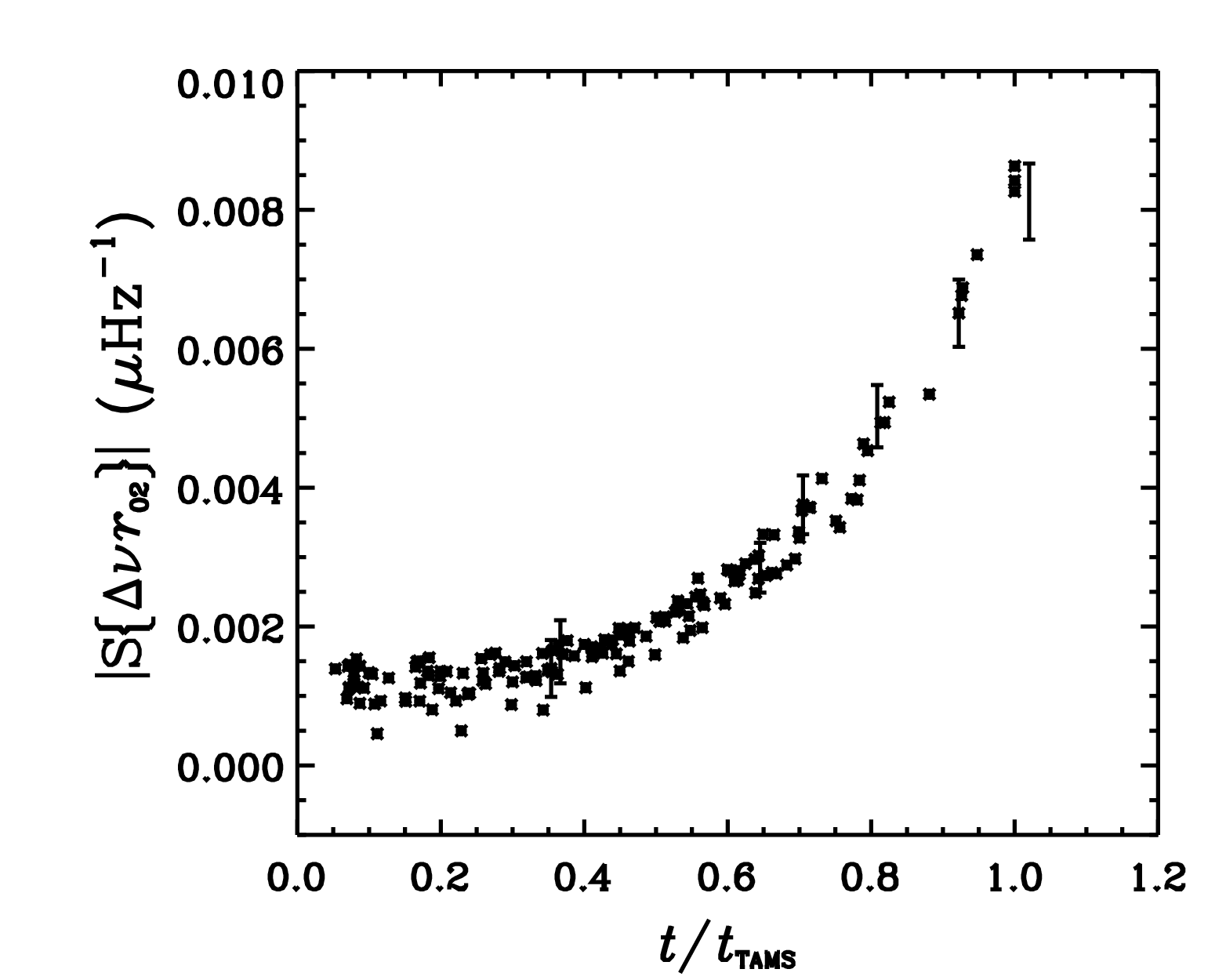} 
\end{center}
\caption{Upper panel: the same as in Fig. \ref{fig:slopedr0213_tttams_all}
but here only considering models with $M \geq 1.3\,\rm M_{\sun}$. The same models
are shown in the middle and lower panels for the slopes of the 
diagnostic tools $\Delta\nu\,r_{010}$ and $\Delta\nu\,r_{02}$, respectively.}
\label{fig:slope_all_tttams}
\end{figure}

One interesting consequence of this overshoot-dependent 
slope dispersion seen in the lower mass models of our grid 
is that it opens the perspective of constraining  
the extent of overshoot present at the border of 
convective regions in stars based on the analysis of the slopes,
if the stellar mass can be accurately estimated.

In fact, for relatively  evolved $M=1.1\rm\,M_{\sun}$ models with solar metallicity, 
the slopes depend strongly on overshoot and a measurement of a 
relatively large absolute value for the slope 
in a star at such mass would imply the presence of significant overshoot. 

Fig. \ref{fig:slope_all_tttams}, upper panel, shows the same plot
as in Fig. \ref{fig:slopedr0213_tttams_all} but for models
with $M \geq 1.3 \rm M_{\sun}$ only. 
The dispersion seen in this case is
significantly smaller than in Fig. \ref{fig:slopedr0213_tttams_all}
because these higher mass models have a relatively large 
convective-core region and, thus, show the expected age$-A_\delta$ relation, even when  $\alpha_{\rm OV}=0$. 
Fig. \ref{fig:slope_all_tttams}, middle and lower panels
show, respectively, the absolute value of the slopes of the diagnostic tools $\Delta\nu\,r_{010}$ and 
$\Delta\nu\,r_{02}$, as a function of the fraction of evolution,
for the models with $M \geq 1.3\,\rm M_{\sun}$. 
From these, we may conclude that
for relatively massive solar-like pulsators 
($M \geq 1.3\,\rm M_{\sun}$), some direct constraints
to the fraction of evolution along the main-sequence
can be derived by inspecting the slopes of the diagnostic tools considered
here, in particular the $\Delta\nu\,dr_{0213}$ and the $\Delta\nu\,r_{010}$.

Such constrains are stronger for relatively evolved stars, 
when the age dependence of the slopes is more accentuated.
However, in this case they are  
limited to metallic stars, since only for those one may observe
the frequency region where the slope is computed.
\section{Conclusions}
\label{sec5}
We performed a systematic study of the slopes of the diagnostic tools $dr_{0213}$, 
$r_{010}$ and $r_{02}$, where the slopes are defined as 
the frequency derivatives of these quantities taken at their 
maximum absolute value. For this study, we considered stellar models 
of different masses, metallicities and convective-core overshoots
and at different evolutionary states in the main sequence.

We verified that for each evolutionary sequence, 
the absolute value of the slopes increases as the 
star evolves on the main sequence. This increase 
is associated with the increase in the sound-speed 
discontinuity at the edge of the core. We determined, 
for each diagnostic tool, the maximum absolute value 
that the slope may take and provided evidence that 
these maxima result from the existence of a 
maximum discontinuity in the mean molecular weight.
The maximum absolute values 
that we found for the slopes of $\Delta\nu\,dr_{0213}$,
$\Delta\nu\,r_{010}$ and $\Delta\nu\,r_{02}$ 
were, respectively, $\sim$0.008, $\sim$0.011 and $\sim$0.009$\,\mu \rm Hz^{-1}$.
Thus, if one such slope is found in a star, 
one can confidently assume that the star is very close to the TAMS.

In addition, we found that models with 
$\left| S\{\Delta\nu\,dr_{0213}\}\right| \gtrsim 0.002$,
$\left| S\{\Delta\nu\,r_{010}\}\right| \gtrsim 0.003\,\mu \rm Hz^{-1}$
and $\left| S\{\Delta\nu\,r_{02}\}\right| \gtrsim 0.0035\,\mu \rm Hz^{-1}$ all have convective cores.
Within these models, those for which the measured slope is in the expected observed
frequency range, namely between $\nu_{\rm max}-8\Delta\nu$ and $\nu_{\rm max}+8\Delta\nu$,
are the most metallic ones, with $Z/X=0.0787$.
Models with $\left| S\{\Delta\nu\,dr_{0213}\}\right| \lesssim 0.002$,
$\left| S\{\Delta\nu\,r_{010}\}\right| \lesssim 0.003\,\mu \rm Hz^{-1}$
and  $\left| S\{\Delta\nu\,r_{02}\}\right| \lesssim 0.003\,\mu \rm Hz^{-1}$
may or may not have a convective core, depending on the mass.

For all diagnostic tools, we found a strong correlation between the
slopes and the relative amplitude $A_\delta$ of the discontinuity in the 
squared sound speed.
We thus conclude that  when the slopes of the diagnostic 
tools can be computed from the observations, 
they provide a direct measurement of the relative 
amplitude of the sound-speed discontinuity at the edge of the core. 
We note, however, that the slopes, as defined in this work, 
can only be computed in practice for stars in which 
the range of observed frequencies contains the frequency 
where the derivative of the diagnostic tools reach their 
maximum absolute value. We have shown that at least 35 per cent
of stars in our sample are expected to satisfy this condition. 

Finally, we considered in more detail the 
relation between the slopes of the 
diagnostic tools and the fraction of stellar 
main-sequence evolution, $t/t_{\rm TAMS}$. 
This relation is 
stronger for the diagnostic tools 
$dr_{0213}$ and $r_{010}$ than for $r_{02}$. 
Also, the dispersion seen in these 
relations is significantly reduced when only models with masses 
$M \geq 1.3\,\rm M_{\sun}$ are considered. 
For these masses, one may thus be able to use
the slopes of $dr_{0213}$ or $r_{010}$ to infer about $t/t_{\rm TAMS}$,
at least for stars in which the slope can be computed from the observations.
On the other hand, for stars with masses $M \leq 1.2\,\rm M_{\sun}$, 
the dispersion seen in the slope versus $t/t_{\rm TAMS}$ relation was found 
to be directly related to the amount of core overshoot, 
which in these lower mass stars, with very small convective cores, 
influences the amplitude of the discontinuity in the mean molecular weight
at fixed evolutionary stage. 
As a consequence, for these lower mass stars, it may be possible to 
use the slopes to discriminate against models with small amounts of core overshoot. 
\section*{Acknowledgments}
IMB acknowledges the support from the Funda\c{c}\~ao para a Ci\^encia e
Tecnologia (Portugal) through the grant SFRH/BPD/87857/2012.
MSC is supported by an Investigador FCT
contract funded by FCT/MCTES (Portugal) and POPH/FSE (EC). IMB and MSC
acknowledge the support from ERC, under FP7/EC, through the project FP7-SPACE-2012-312844.
Funding for the Stellar Astrophysics Centre is provided by The Danish National Research
Foundation (Grant DNRF106). The research is supported by the ASTERISK project
(ASTERoseismic Investigations with SONG and \textit{Kepler}) funded by the European Research
Council (grant agreement no.: 267864).
\bibliographystyle{mn2e}
\bibliography{core_v1.bib} 
\label{lastpage}
\end{document}